\documentclass[acmtog]{acmart}
\usepackage{multirow}
\usepackage[ruled]{algorithm2e} %

\SetAlFnt{\small}
\SetAlCapFnt{\small}
\SetAlCapNameFnt{\small}
\SetAlCapHSkip{0pt}

\usepackage[capitalize]{cleveref}
\usepackage{placeins}

\copyrightyear{2025}
\acmYear{2025}
\setcopyright{cc}
\setcctype{by}
\acmConference[SIGGRAPH Conference Papers '25]{Special Interest Group on Computer Graphics and Interactive Techniques Conference Conference Papers }{August 10--14, 2025}{Vancouver, BC, Canada}
\acmBooktitle{Special Interest Group on Computer Graphics and Interactive Techniques Conference Conference Papers (SIGGRAPH Conference Papers '25), August 10--14, 2025, Vancouver, BC, Canada}
\acmDOI{10.1145/3721238.3730701}
\acmISBN{979-8-4007-1540-2/2025/08}

\def\MethodName{Elevate3D} 
\def\RefineMethodName{HFS-SDEdit}

\NewEnviron{redblock}{%
  \color{red}%
  \BODY%
} %

\graphicspath{{./figures/}}

\begin{document}
\title{Elevating 3D Models: High-Quality Texture and Geometry Refinement from a Low-Quality Model}

\author{Nuri Ryu}
\orcid{0000-0002-7769-689X}
\affiliation{%
  \institution{POSTECH}
   \country{South Korea}
}
\email{ryunuri@postech.ac.kr}

\author{Jiyun Won}
\orcid{0009-0003-6233-0988}
\affiliation{%
  \institution{POSTECH}
   \country{South Korea}
}
\email{w1jyun@postech.ac.kr}

\author{Jooeun Son}
\orcid{0009-0003-6916-9912}
\affiliation{%
  \institution{POSTECH}
   \country{South Korea}
}
\email{jeson@postech.ac.kr}

\author{Minsu Gong}
\orcid{0009-0008-9304-5203}
\affiliation{%
  \institution{POSTECH}
   \country{South Korea}
}
\email{gongms@postech.ac.kr}

\author{Joo-Haeng Lee}
\orcid{0000-0002-5788-712X}
\affiliation{%
  \institution{Pebblous}
   \country{South Korea}
}
\email{joohaeng@pebblous.ai}

\author{Sunghyun Cho}
\orcid{0000-0001-7627-3513}
\affiliation{%
  \institution{POSTECH}
   \country{South Korea}
}
\email{s.cho@postech.ac.kr}

\renewcommand{\shortauthors}{Ryu et al.}

\begin{teaserfigure}
    \centering
\includegraphics[width=\linewidth,trim=0cm 0.5cm 0.cm 0.5cm]
    {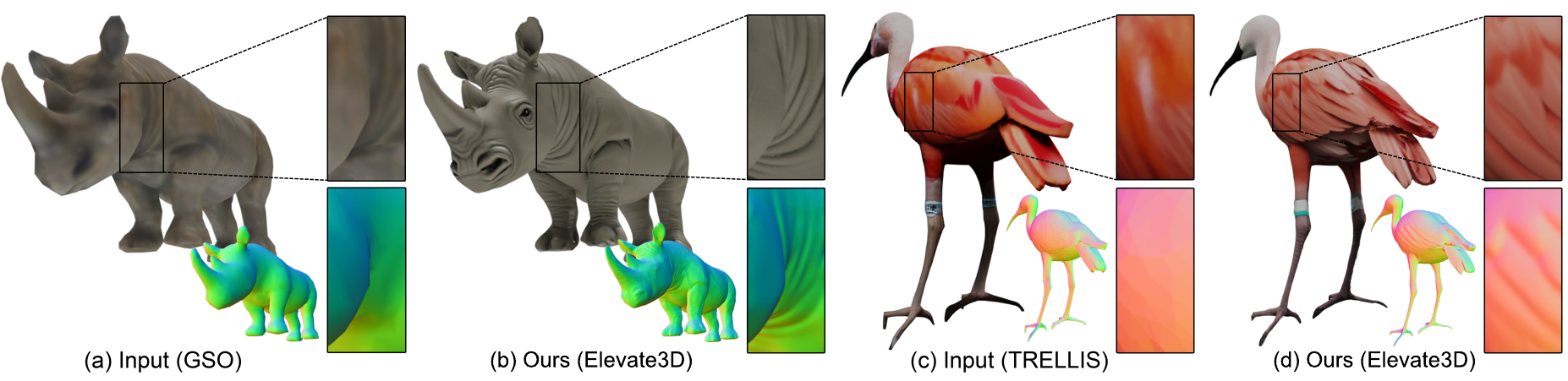}
    \caption{
    3D refinement examples from (a) a degraded real-world scan \cite{GSO} and (c) a state-of-the-art image-to-3D generative model \cite{trellis}. Our method, \MethodName{}, effectively refines both texture and geometry while preserving their alignment, as shown in (b) and (d). Inputs for the experiment: the GSO dataset~\cite{GSO}, and ©MasaStojanovic/pixabay.
    }
    \label{fig:teaser}
\end{teaserfigure}

\begin{abstract}
High-quality 3D assets are essential for various applications in computer graphics and 3D vision but remain scarce due to significant acquisition costs. To address this shortage, we introduce \MethodName{}, a novel framework that transforms readily accessible low-quality 3D assets into higher quality. At the core of \MethodName{} is \RefineMethodName{}, a specialized texture enhancement method that significantly improves texture quality while preserving the appearance and geometry while fixing its degradations. Furthermore, \MethodName{} operates in a view-by-view manner, alternating between texture and geometry refinement. Unlike previous methods that have largely overlooked geometry refinement, our framework leverages geometric cues from images refined with \RefineMethodName{}
by employing state-of-the-art monocular geometry predictors. 
This approach ensures detailed and accurate geometry that aligns seamlessly with the enhanced texture. \MethodName{} outperforms recent competitors by achieving state-of-the-art quality in 3D model refinement, effectively addressing the scarcity of high-quality open-source 3D assets.
\end{abstract}



\maketitle

\section{Introduction}
\label{sec:intro}

High-quality 3D models are in unprecedented demand, serving as essential components for various applications and as training data in computer graphics and 3D vision~\cite{euro_star_neural_rendering, deepgenerativemodels3dsurvey, 3Dseg_survey}.
Despite the growth in large-scale open-source 3D model datasets~\cite{objaverse, Objaverse-XL, omniobject3d}, high-quality models remain scarce due to high acquisition costs.
To address this, we tackle the problem of texture and geometry refinement of easily accessible low-quality models, bridging the gap between abundant low-quality data and the pressing need for high-quality models.

Constructing high-quality 3D models from low-quality counterparts, a process we refer to as 3D model refinement, is of great importance but has received relatively less attention.
Traditional methods, such as mesh subdivision and denoising, primarily focus on refining geometry by suppressing noisy structures or smoothing surfaces using simple geometric priors. However, these methods fall short of producing high-quality geometric details as they rely solely on simple priors and do not address texture refinement.

Recently, several 3D model refinement methods have emerged, either as post-processing steps within 3D model generation pipelines or as stand-alone methods for refining existing low-quality models~\cite{DreamCraft3D, dreamgaussian, 
magicboost, repaint123, unique3d, CLAY, disr}. These approaches leverage generative priors, such as single-image and multi-view diffusion models~\cite{dickstein_ddpm, DDPM} and GANs~\cite{goodfellow2014generative}, to enhance both texture and geometry or texture alone. Specifically, they render multiple views of a low-quality model, enhance each view, and update the 3D model.
One common strategy for view enhancement is to apply diffusion- or GAN-based image enhancement models specifically trained to refine the appearance of 3D models~\cite{magicboost, unique3d, CLAY}. Another popular approach is to employ SDEdit~\cite{sdedit}, which offers high flexibility, enabling refinement of low-quality models with various degradations~\cite{DreamCraft3D, dreamgaussian, 
magicboost, repaint123}.

Despite the advances in 3D model refinement, limitations in both texture and geometry still remain. Most recent approaches refine each view independently, leading to cross-view inconsistencies and blurry textures. While multi-view diffusion models may improve consistency, they are limited in image resolution due to large memory requirements, restricting refinement quality~\cite{wonder3d}.

SDEdit-based approaches also suffer from a trade-off between fidelity to the input and the quality of refined 3D models. SDEdit generates structured noise by interpolating the input image with Gaussian noise, and applies the diffusion model’s denoising process~\cite{dickstein_ddpm, DDPM}. This aligns the image with the high-quality distribution learned by the diffusion model while preserving key features of the input. However, this process imposes a quality-fidelity trade-off based on the noise level. Stronger noise enhances conformance to the diffusion model's distribution but harms fidelity. In contrast, lower noise maintains fidelity but offers only minor enhancements.

Lastly, previous approaches rely solely on image-based priors derived from large-scale image generative models.
This exclusive reliance on photometric constraints leaves the geometry under-constrained; even if the refined models appear visually appealing from certain viewpoints, their underlying geometric structure may still be of poor quality~\cite{MonoSDF, mip_nerf_360, nerf_pp}.
To address these issues, some approaches incorporate separate geometry and texture optimization stages~\cite{DreamCraft3D, dreamgaussian}. However, decoupling the two processes often results in misaligned texture and geometry~\cite{SiTH}.

This paper proposes \emph{\MethodName{}}, a novel 3D model refinement approach that produces a high-quality 3D model with well-aligned texture and geometry. \MethodName{} operates iteratively, refining textures and geometries view-by-view. For each view, it first enhances the texture and then leverages the refined texture to adjust the geometry, ensuring alignment. In subsequent views, the texture refinement is based on the visible parts of the updated geometry and the original geometry. This process ensures precise alignment between texture and geometry and allows the use of high-quality priors trained on large-scale image datasets for both texture and geometry, ultimately producing superior 3D models.

For texture refinement, \MethodName{} adopts \textit{High-frequency-Swapping SDEdit} (\RefineMethodName{}), a specialized method that resolves the fidelity–quality trade-off in SDEdit.
Our key insight leverages the coarse-to-fine nature of the diffusion model’s generative process~\cite{generative_heat}, where low-frequency features are established first and strongly influence subsequent high-frequency detail generation. 
If these low-frequency features are constrained to match a low-quality input, the final image inevitably inherits unwanted artifacts.
Instead, we let the diffusion model freely generate low-frequency features while applying constraints only high-frequency components to match the input image during the early denoising steps.
In this way, the generative path is steered toward the high-quality image distribution while minimal high-frequency guidance preserves crucial edges and details—sufficient to maintain the input’s unique identity without embedding its artifacts. 

After refining the texture at a viewpoint, we enhance the geometry for that viewpoint using the refined texture. We employ a state-of-the-art monocular normal predictor~\cite{mari_e2e} to infer detailed surface normals aligned with the updated texture.
The predicted normals may not perfectly match the initial 3D geometry. Thus, we employ a regularized normal integration scheme to estimate a geometry consistent with the initial 3D geometry from the predicted normals.
We then stitch the estimated geometry with previously refined or unrefined regions, maintaining consistent geometry across viewpoints.
This process produces a  3D representation faithful to the refined textures and supports direct texture projection onto the enhanced geometry without misalignment.

Comprehensive experiments demonstrate that \MethodName{} successfully produces high-quality 3D models from low-quality ones, including those generated by previous 3D generation models and low-quality scanned models. 
To summarize, our main contributions are as follows:
\begin{itemize}
    \item We propose a novel 3D model refinement framework, \MethodName{}, which alternates between texture and geometry refinement in a view-by-view fashion to produce a high-quality 3D model with well-aligned texture and geometry
    \item We introduce \RefineMethodName{} for texture refinement, leveraging high-frequency guidance to achieve high-quality and high-fidelity enhancements while mitigating the limitations of previous SDEdit-based methods. 
    \item We demonstrate that our framework can achieve state-of-the-art quality refinement of 3D models compared with recent competitors through comprehensive experiments.
\end{itemize}

\section{Related Work}
\label{sec:related}

\paragraph{3D Model Generation}
Recent advances in 3D model generation leverage diffusion models due to their powerful image priors~\cite{3dgenerationsurvey}. Initially, the SDS loss was introduced to use pre-trained text-to-image diffusion models for generating 3D models directly from text or image prompts~\cite{DreamFusion, neuralLift, makeit3d, realfusion, prolificdreamer}. Although promising, these methods often produce textures with over-saturated colors and blurred details.
Alongside these optimization approaches, early efforts also adapted 2D diffusion models to generate multi-view consistent novel views via inpainting for 3D reconstruction~\cite{POP3D, 10.1145/3610548.3618149_iNVS}.
These were succeeded by a line of works that fine-tune pre-trained diffusion models to generate multi-view images for 3D reconstruction~\cite{wonder3d, syncdreamer, mvdream, zero123plus}. However, these models require additional components on top of the large-scale diffusion model~\cite{StableDiffusion} to enforce multi-view consistency. This added complexity increases computational costs and memory requirements, limiting both the number of views that can be generated simultaneously and their resolution, consequently affecting the quality of the generated 3D models~\cite{wonder3d}.
In our work, we demonstrate that our refinement method can successfully produce high-quality 3D models from low-quality ones generated by previous approaches.

\paragraph{3D Model Refinement}

Recently, 3D model refinement methods have primarily emerged as components of 3D model generation pipelines.
To improve the low-quality outputs from a coarse 3D generation step, recent pipelines often introduce a relatively straightforward refinement stage.
One widely adopted approach is to use SDEdit~\cite{sdedit} with an image-based loss~\cite{DreamCraft3D, dreamgaussian, magicboost, repaint123}. 
Specifically, a low-quality image is rendered from a coarse 3D model at an arbitrary viewpoint and then refined using SDEdit. The 3D model is updated using the refined images through an MSE loss.
However, this approach has significant drawbacks. First, refined details at each view are not multi-view consistent; details introduced in one viewpoint are averaged out across others, resulting in blurry outcomes. Second, the image-based loss under-constrains the geometry, forcing such methods to keep the geometry fixed and focus solely on texture refinement.
Another line of approaches~\cite{unique3d,CLAY} generates multi-view images and refines them using off-the-shelf super-resolution networks such as Real-ESRGAN~\cite{realesrgan}. The refined images are then used to synthesize 3D models. However, these methods perform super-resolution on each image independently, resulting in blurriness due to multi-view inconsistency.

Standalone 3D refinement methods outside of 3D generation pipelines encounter similar challenges. For instance, MagicBoost~\cite{magicboost} employs multi-view diffusion models combined with SDS optimization to enhance both geometry and texture. However, artifacts introduced during SDS optimization necessitate additional refinement using SDEdit, leading to the same multi-view consistency issues. DiSR-NeRF~\cite{disr} pairs an SDS loss~\cite{DreamFusion} with a diffusion-based 2D super-resolution model~\cite{StableDiffusion}. Although this method iteratively refines a low-resolution NeRF by enhancing 2D images and synchronizing the 3D model, it prioritizes consistency in low-resolution features. Consequently, high-resolution details generated during refinement can still be averaged out during 3D synchronization.

In contrast, \MethodName{} adopts a novel approach to refinement by preserving and distinguishing previously enhanced regions in each view. This ensures a multi-view consistent result across all viewpoints. Additionally, by alternating between texture and geometry refinement steps, \MethodName{} allows the geometry to be refined according to the improved textures. This strategy addresses the limitations of earlier methods, which either average out high-frequency details or neglect geometry refinement altogether.

\paragraph{3D Model Texturing}
Another relevant task related to ours is 3D model texturing~\cite{Text2Tex, TEXTure, intex, paint3d, paint-it}. Texturing approaches synthesize textures for 3D models without altering the geometry. Hence, these methods, like 3D model refinement approaches, face the limitation of texture-geometry misalignment. Even if an input 3D model lacks geometric details, these methods may still produce textures with rich details learned from generative priors, leading to inconsistencies between texture and geometry. This highlights the need for methods that jointly refine both texture and geometry. Our proposed \MethodName{} addresses the shortcomings of one-sided methods by integrating both aspects within a unified framework.
\begin{figure*}[t!]    
\begin{center}
\includegraphics
[width=\linewidth, trim=0cm 0cm 0cm 0cm]
{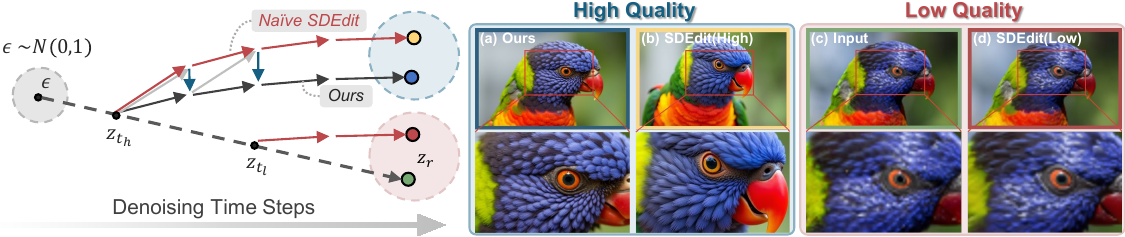}
\end{center}
\caption{\textbf{Overview of \RefineMethodName{}.} \RefineMethodName{} addresses the quality-fidelity trade-off in SDEdit. By adding a substantial amount of noise $\epsilon$ to the low-quality reference image $z_r$ in (c) and initiating the denoising process from the noisy latent $z_{t_\text{h}}$, SDEdit removes domain information, enabling the diffusion model to generate a high-quality image as depicted in (b). However, this approach compromises fidelity to the reference image. Conversely, adding a small amount of noise and starting the denoising process from the noisy latent $z_{t_\text{l}}$ preserves the low-quality domain information, resulting in only minor refinements, as seen in (d). In contrast, \RefineMethodName{} incorporates high-frequency feature injection-based guidance, as detailed in \cref{sec:HFSDiff}, allowing for high-fidelity generation even when starting the denoising process from $z_{t_\text{h}}$. This approach achieves both high quality and high fidelity in the refinement process. Input: ©Watts/flickr.}
\label{fig:method_diagram}
\end{figure*}

\section{\RefineMethodName{} for Image Refinement}
\label{sec:HFSDiff}

\RefineMethodName{}, built on top of SDEdit~\cite{sdedit}, is a critical component of \MethodName{} for high-quality, high-fidelity texture refinement. This section reviews SDEdit then introduces \RefineMethodName{}.

\paragraph{SDEdit}
SDEdit is a technique designed to guide the image synthesis process of diffusion models. Specifically, for a given reference image, such as a stroke image or a low-quality image, the goal of SDEdit is to generate a realistic image that aligns with the learned distribution of the diffusion model and adheres to the structure of the reference image.
SDEdit leverages the observation that adding more noise to images in both the reference image domain ($\mathcal{I}_r$) and the realistic image domain ($\mathcal{I}$) gradually merges them, transforming them into the domain of Gaussian noise.
Based on this, SDEdit proposes a simple, training-free strategy.
Let $z_r \in \mathcal{I}_r$ represent a reference image in the latent space of a diffusion model.
SDEdit then initalizes a noisy latent $z_{t_s}$ by adding noise to $z_r$ as:
\begin{equation}
    z_{t_s} = \alpha(t_s)z_r + \beta(t_s)\epsilon,
    \label{eq:sdedit}
\end{equation}
where $t_s$ is a timestep in the denoising schedule of the diffusion model such that $t_s\in\{T,T-1,\cdots,0\}$. The functions $\alpha(t)\in[0,1]$ and $\beta(t)\in[0,1]$ control the noise level at timestep $t$.
Starting from this initial noisy latent, SDEdit performs the iterative denoising process to produce a realistic image $z_0$.

SDEdit offers several distinct benefits, including easy integration into diffusion-based image synthesis pipelines, training-free implementation, and computational efficiency. Thus, it has been widely adopted for various tasks, e.g., image editing~\cite{sdedit_image_edit}, video generation~\cite{sdedit_video}, 3D editing~\cite{MVEdit}. However, it exhibits a fidelity-quality trade-off, which will be discussed in detail later.

\paragraph{\RefineMethodName{}}
To overcome the fidelity-quality trade-off of SDEdit, our proposed \RefineMethodName{} replaces the high-frequency component of the latent representation with that of the reference image to ensure that the synthesized image $z_0$ aligns with the structural details of the reference image $z_r$.
Specifically, \RefineMethodName{} starts with \cref{eq:sdedit}.
Then, in the subsequent timesteps $t$, \RefineMethodName{} replaces the high-frequency component of the latent $\hat{z}_t$, the output of the standard denoising process, by evaluating:
\begin{align}
    \tilde{z}_{t} &= \alpha(t)z_r + \beta(t)\epsilon,\qquad\qquad\text{and}\label{eq:hfs1}\\
    z'_t &= (\delta-G_\sigma)*\tilde{z}_t + G_\sigma*\hat{z}_t
    \label{eq:hfs2}
\end{align}
where $\tilde{z}_t$ is a noised reference image and $z'_t$ is a calibrated latent whose high-frequency component is replaced with that of $\tilde{z}_t$. $\delta$ is a Dirac delta function, $G_\sigma$ is a Guassian kernel with standard deviation $\sigma$, and $*$ is the convolution operator. Once $z'_t$ is obtained, denoising is performed with $z'_t$ resulting in $\hat{z}_{t-1}$.
We perform high-frequency replacement until $t$ reaches a predefined timestep $t_\text{stop}$ to ensure that the resulting image does not reproduce all the details, such as the degraded details of the low-quality reference image.

\RefineMethodName{} is based on the intuition that the \emph{domain} information of an image is encoded in the mid-frequency component of the latent representation.
Specifically, we may decompose the latent representation into two frequency bands: low-, and high-frequency components. Similar to conventional images, the high-frequency component describes small-scale structures~\cite{understanding_latent_space}. On the other hand, the low-frequency component contains not only large-scale structures but also domain information.

\cref{fig:method_diagram} illustrates the intuition behind \RefineMethodName{} as well as the quality-fidelity trade-off of SDEdit with an image enhancement example. Adding more noise to the reference image $z_r$ gradually removes information from high-frequency to low-frequency components. Thus, to sufficiently merge the realistic image domain and the reference image domain, or equivalently to remove domain information, SDEdit requires the use of large noise to remove a sufficient amount of low-frequency components.
However, such excessive noise also destroys small- and large-scale structures, causing high-quality but low-fidelity results.
Conversely, SDEdit with small noise removes only the high-frequency component, producing an image not only with different small-scale details but also within the same domain as the reference image, the low-quality image domain.

To improve both quality and fidelity, \RefineMethodName{} starts with large noise and injects high-frequency components of the reference image into the synthesis process.
This approach effectively removes the domain information of the reference image, resulting in a high-quality image.
The injected high-frequency component not only constrains the high-frequency details of a synthesized image, but also guides the diffusion model to synthesize a low-frequency component that aligns with the injected details, achieving high-fidelity synthesis.

\begin{figure}[t!]    
\begin{center}
\includegraphics[width=\columnwidth,trim=0cm 0cm 0cm 0cm]{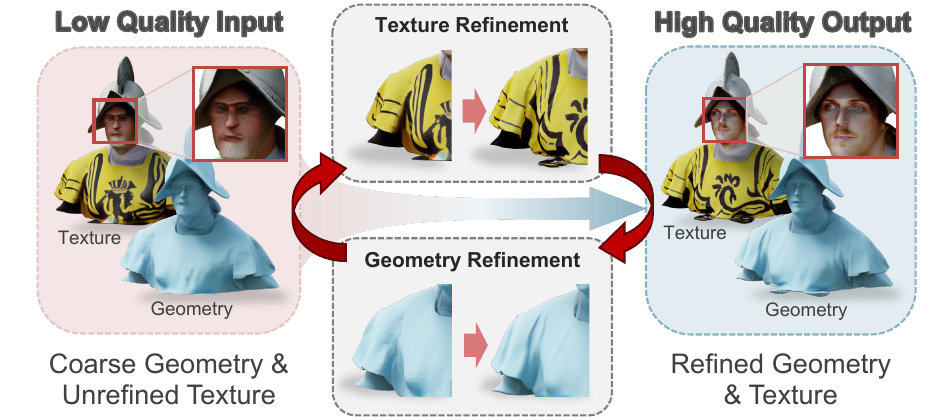}
\end{center}
\caption{\textbf{Framework Overview.} Given a low-quality 3D model, \MethodName{} alternatingly refines texture and geometry. Input for the experiment: ©Momentmal/pixabay.
}
\label{fig:framework_overview}
\end{figure}

\section{\MethodName{}}
\label{sec:method}

\cref{fig:framework_overview} visualizes an overview of \MethodName{}. Given a low-quality textured mesh, our approach progressively refines both the texture and geometry of the mesh by iterating through a predefined camera path $\mathcal{V}=\{v_0, \ldots, v_k\}$. Specifically, at the $i$-th iteration corresponding to $v_i$, we denote the initial partially refined model as $M_i$, where $M_0$ is initialized as the input low-quality textured mesh. Our method then refines the unrefined regions of the texture and geometry of $M_i$ that are visible at $v_i$ through texture refinement and geometry refinement stages. The texture refinement stage focuses on enhancing the unrefined regions while maintaining consistency with the already refined areas. Afterward, the geometry refinement stage extracts geometric cues from the newly refined texture and refines the mesh geometry accordingly. This ensures that the geometry accurately reflects the details present in the refined texture, maintaining texture-geometry consistency. 

This view-by-view refinement strategy enables our method to utilize predefined image and geometry priors, which contribute to achieving high-quality texture and geometry. Additionally, by leveraging the refined texture and geometry when processing unrefined regions in subsequent viewpoints, our method ensures cross-view consistency. Also, the geometry refinement is performed based on the refined texture, significantly enhancing texture-geometry consistency. In the following, we explain each stage in detail.

\subsection{Texture Refinement with \RefineMethodName{}}
\label{subsec:\RefineMethodName{}}

\begin{figure}[t!]    
\begin{center}
\includegraphics[width=\linewidth, trim=0cm 0cm 0cm 0cm]{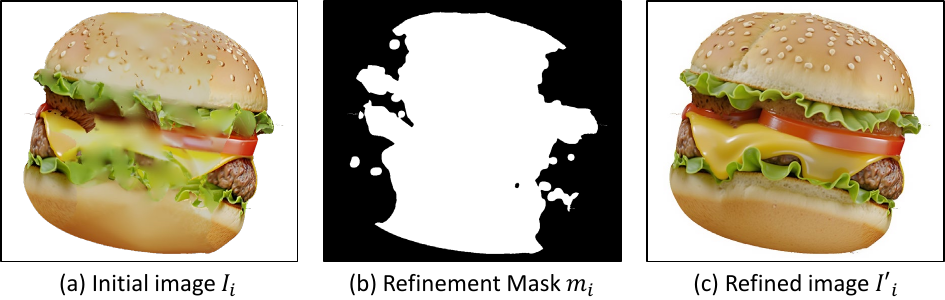}
\end{center}
\caption{\textbf{Texture Refinement.}
Given a partially-refined texture image $I_i$ in (a), 
the texture refinement stage detects a refinement mask $m_i$ in (b), and produces a refined image in (c) using \RefineMethodName{}.}
\label{fig:method_texture_refine}
\end{figure}

The texture refinement stage improves the texture of the partially refined 3D model $M_i$ for the current view $v_i$ by leveraging \RefineMethodName{} with a large-scale pretrained image diffusion model. We begin by rendering $M_i$ from $v_i$ to obtain the initial image $I_i$ (\cref{fig:method_texture_refine}-a). This image contains regions refined at previous views $\{v_0, \dots, v_{i-1}\}$ and unrefined regions that still exhibit low-quality textures.

To isolate the unrefined regions in $I_i$, we identify pixels that are not visible in previous views. Specifically, we rasterize normal vectors for $v_i$ and compare them with the viewing directions of the previous views $\{v_0, \dots, v_{i-1}\}$ by computing cosine similarities. We detect pixels whose cosine similarity values exceed a threshold $\tau$ for all previous views and construct a binary mask $m_i$ (\cref{fig:method_texture_refine}-b). We set $\tau=0.5$, equivalent to a $60^\circ$ angle difference. This approach may include already-refined pixels in the mask, but it allows regions seen at oblique angles to be further refined.

Once $m_i$ is obtained, we refine the unrefined regions in $I_i$ using \RefineMethodName{}. To refine only the unrefined regions specified by $m_i$, we introduce a slight modification to \RefineMethodName{}. Specifically, at each timestep of the diffusion sampling process, we first compute a noised reference $\tilde{z}_t$ and a latent $z'_t$ using \cref{eq:hfs1,eq:hfs2}. To ensure that the regions outside $m_i$ are not overwritten, we blend $z'_t$ with $\tilde{z}_t$ using $\tilde{m}$, a downsampled version of $m_i$. This blending is done as:
\begin{equation}
    \label{eq:blended}
    \hat{\mathbf{z}}_t = \tilde{\mathbf{m}} \odot \mathbf{z}'_t + \bigl(1 - \tilde{\mathbf{m}}\bigr) \odot \tilde{\mathbf{z}}_t.
\end{equation}
Finally, $\hat{\mathbf{z}}_t$ goes through the denoising step. After iterative denoising with \RefineMethodName{}, we obtain a refined image $I'_i$ with improved textures faithfully reflecting $I_i$ for the unrefined regions while preserving the textures from $I_i$ in the already-refined regions (\cref{fig:method_texture_refine}-c).

\subsection{Geometry Refinement with Refined Texture}
\label{subsec:geo_refine}

The geometry refinement step enhances the geometry of the partially refined 3D triangle mesh \( M_i \) by utilizing the refined texture image \( I'_i \) from the previous texture refinement stage. 
This image not only exhibits improved textures but also provides valuable cues for geometric details. 
These details can be extracted using monocular geometry estimation models~\cite{marigold, mari_e2e, depth_anything_v2}.

In \MethodName{}, we start by inferring a normal map \( \mathbf{n}_i \) from the refined image \( I'_i \) using a state-of-the-art normal estimation model~\cite{mari_e2e}.
We then integrate the estimated normals \( \mathbf{n}_i \) to obtain a refined surface \( S_i \) corresponding to the viewpoint \( v_i \). This surface is consistent with the refined texture image \( I' \).
However, because \( \mathbf{n}_i \) is derived solely from \( I'_i \) and not conditioned on the existing geometry of \( M_i \), the refined geometry \( S_i \) can significantly deviate from that of \( M_i \). 
Consequently, directly stitching \( S_i \) onto \( M_i \) could introduce severe geometric distortion.

To address this, we introduce a regularized normal integration scheme to estimate \( S_i \) while ensuring consistency with the existing geometry of \( M_i \).
Our regularized normal integration scheme is implemented as follows.
We assume an orthographic camera model and rasterize a depth map \( d \) of \( M_i \) from viewpoint \( v_i \).
We define \( S_i \) and \( \mathbf{n}_i \) as \( S_i(u,v)=[u,v,z(u,v)]^\top \), and \( \mathbf{n}_i(u,v)=[n_x(u,v), n_y(u,v), n_z(u,v)]^\top \), respectively, where \( u \) and \( v \) represent pixel coordinates.
Our goal is to estimate the depth component \( z(u,v) \)  such that the resulting surface \( S_i \) satisfies the following two criteria:
First, the normals of \( S_i \) must be close to \( \mathbf{n}_i \), or equivalently, its tangent vectors \( \partial S_i(u,v) / \partial u \) and \( \partial S_i(u,v) / \partial v \) are orthogonal to \( \mathbf{n}_i(u,v) \).
Second, the estimated depth \( z(u,v) \) must be close to the depth \( d(u,v) \) rendered from \( M_i \).
Based on these two criteria, we define an energy functional:
\begin{equation}\label{eq:energy}
\begin{split}
E(z) = \iint \Bigl[\, 
&\Bigl(\frac{\partial z}{\partial u} + \frac{n_x(u,v)}{n_z(u,v)}\Bigr)^2 
+ \Bigl(\frac{\partial z}{\partial v} + \frac{n_y(u,v)}{n_z(u,v)}\Bigr)^2 
\Bigr] \,du\,dv \\
&\quad + \lambda \iint \Bigl(z(u,v) - d(u,v)\Bigr)^2 \,du\,dv,
\end{split}
\end{equation}
where the first and second terms on the right-hand-side correspond to the first and second criteria, respectively.
\( \lambda \) is a regularization parameter that balances the two terms. Minimizing \( E(z) \) yields a refined depth map for \( S_i \). In our implementation, we follow the normal integration method of Cao et al.~\shortcite{bini} to minimize \( E(z) \).

Once we obtain the refined geometry \( S_i \) represented by the refined depth map, we update the mesh \( M_i \). Specifically, the refined geometry region \( S_i \) is integrated with the unchanged parts of \( M_i \) using Poisson surface reconstruction~\cite{poisson_surface}. 
This step generates the updated triangular mesh \( \tilde{M}_i \). While Poisson reconstruction does not strictly preserve the original mesh topology, it rarely introduces geometric artifacts in our case. This robustness stems from our regularized integration in ~\cref{eq:energy}, which constrains the geometric update using the coarse input mesh \( M_i \) via the depth map \( d \) as a guide. This process ensures that each view’s improved geometry is seamlessly integrated without disrupting previously refined areas of the model. 
Furthermore, thanks to the geometry refinement process leveraging the refined texture, \MethodName{} ensures proper texture-geometry alignment.

\begin{figure}[t!]    
\begin{center}
\includegraphics[width=\linewidth]{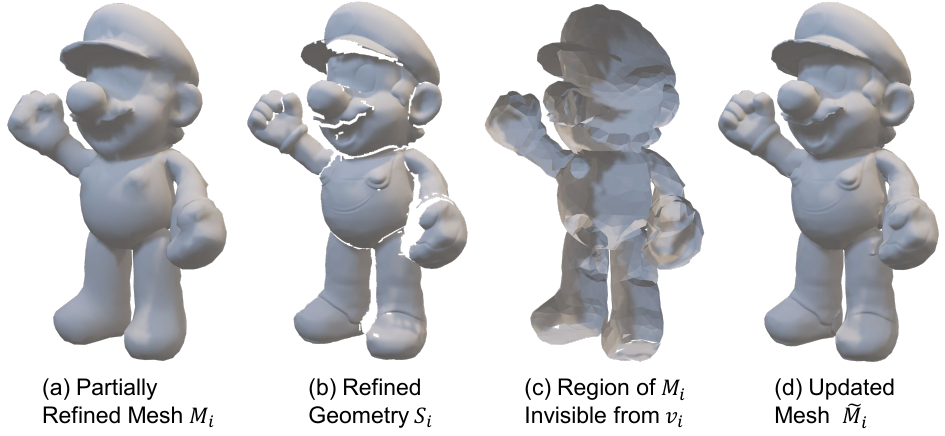}
\end{center}
\caption{\textbf{Geometry Refinement.} Given a partially refined geometry $M_i$ in (a), we obtain a refined surface $S_i$ in (b), and stitch it with the other regions of $M_i$ shown in (c), resulting in the updated mesh $\tilde{M}_i$ in (d). Input for the experiment: the GSO dataset~\cite{GSO}}
\label{fig:method_geometry_refine}
\end{figure}

\cref{fig:method_geometry_refine} illustrates this geometry refinement process.
Given a partially refined mesh \( M_i \) (\cref{fig:method_geometry_refine}-a), our geometry refinement stage estimates a refined geometry \( S_i \), visualized via its depth map (\cref{fig:method_geometry_refine}-b). This refined region is then merged with the other regions of \( M_i \) using Poisson reconstruction (\cref{fig:method_geometry_refine}-c), resulting in the updated mesh \( \tilde{M}_i \) (\cref{fig:method_geometry_refine}-d). We note that the seams in \cref{fig:method_geometry_refine}-b are the result of filtering of unreliable depth values around discontinuities, which we provide details in the supplementary material.

After the geometry refinement stage, we project the refined texture image \( I'_i \) onto the updated mesh \( \tilde{M}_i \) to obtain the textured mesh \( M_{i+1} \) for the next view refinement iteration. 
For this, we employ projection mapping, a form of UV-free texture mapping, similar to recent mesh texturing methods~\cite{Text2Tex, TEXTure, intex}.

\section{Experiments}
\subsection{Implementation Details}
In all experiments, we use FLUX\footnote{\url{https://github.com/black-forest-labs/flux}}---an open-source large-scale text-to-image diffusion model trained with the rectified flow-matching formulation~\cite{liu2022flow, lipman2023flow, albergo2022flow}. We follow the default parameters and the denoising schedule used by FLUX. 
For all experiments, we use a total of $T=30$ denoising steps. 
We set the initial noise timestep $t_s$ to $29$ and the frequency swapping threshold $t_{\text{stop}}$ to $18$. 
We employ a Gaussian low-pass filter with $\sigma=4$. 
These parameters were selected by qualitatively comparing different combinations and choosing the setting that yielded the best qualitative results. 
A quantitative comparison for various combinations of $\sigma$ and $t_{\text{stop}}$ is provided in the supplementary material.
For the depth regularization, we set $\lambda=0.008$. We use an orthographic camera, and for selecting the camera schedule, we adopt a strategy similar to Text2Tex~\cite{Text2Tex}: starting with a pre-defined set of camera poses $\mathcal{V}$, we use an automatic view selection scheme to choose the refinement view that covers the largest unrefined region. We leave the specific details in the supplementary.

\subsection{Evaluation of \MethodName{}}
\label{subsec:3D_exp}
We evaluate the 3D model refinement quality of \MethodName{} using a real-world scan dataset, GSO~\cite{GSO}, where we formed a test set comprising 59 objects. To this end, we degrade the 3D models from the GSO dataset by reducing the number of faces to 20\% and applying a Gaussian low-pass filter with $\sigma=8$ to the textures. We then compare our method with recent 3D model refinement approaches: MagicBoost~\cite{magicboost}, DiSR-NeRF~\cite{disr}, and DreamGaussian~\cite{dreamgaussian}. DreamGaussian focuses solely on texture refinement, while the others refine both texture and geometry. Using each method, we refine the degraded 3D models and compare the quality of the refined models.

\begin{table}[t!]
    \centering
    \caption{\textbf{Quantitative Comparison on 3D Refinement}. \MethodName{} consistently achieves the best scores across various non-reference quality metrics, highlighting its high-quality 3D refinement. DreamGaussian~\cite{dreamgaussian} refines only textures while others refine both texture and geometry. All refinement time were measured using an NVIDIA RTX A6000 GPU.
    }
    \scalebox{0.8}{
        \begin{tabular}{l c c c c c}
            \toprule[1.2pt]
            Model & MUSIQ $\uparrow$ & LIQE $\uparrow$ & TOPIQ $\uparrow$ & Q-Align $\uparrow$ & Time $\downarrow$ \\ 
            \midrule[1.0pt]
            DreamGaussian~\shortcite{dreamgaussian} & 61.667 & 2.1185 & 0.4690 & 2.7416 & \textbf{1 min} \\
            DiSR-NeRF~\shortcite{disr}         & 48.940 & 1.2869 & 0.3879 & 2.6794 & 6 hrs \\
            MagicBoost~\shortcite{magicboost}  & 51.646 & 2.1085 & 0.3915 & 2.4992 & 20 min \\
            \MethodName{} (Ours)              & \textbf{66.527} & \textbf{2.7744} & \textbf{0.5295} & \textbf{3.2151} & 25 min \\
            \bottomrule[1.2pt]
        \end{tabular}
    }    
    \label{table:nvs_cmp}
\end{table}

\cref{fig:qual_comp} shows a qualitative comparison.
As the figure shows, the results of previous approaches show blurry textures and less accurate geometries that are inconsistent with the textures.
In contrast, our method successfully refines input low-quality 3D models, producing detailed textures and geometries, substantially surpassing previous methods. Our results also exhibit high texture-geometry consistency, thanks to our geometry refinement strategy that leverages refined textures.
We also report a quantitative comparison of the quality of the refined 3D models in \cref{table:nvs_cmp}.
For quantitative evaluation, we render the refined models and assess the quality of the rendered images using various image quality metrics: MUSIQ~\cite{MUSIQ}, LIQE~\cite{LIQE}, TOPIQ~\cite{TOPIQ}, and Q-Align~\cite{QAlign}.
As reported in the table, our method consistently outperforms other approaches across all quality metrics.
Regarding computation times,the unoptimized prototype implementation of \MethodName{} operates slower than DreamGaussian, which exclusively refines textures. However, it achieves similar efficiency to MagicBoost and is considerably faster than DiSR-NeRF.

\MethodName{} can also be utilized to produce superior 3D models when combined with state-of-the-art (SoTA) image/text-to-3D synthesis methods. \cref{fig:qual_trellis} illustrates the refinement of 3D models generated by TRELLIS~\cite{trellis}, a leading 3D model synthesis method. A common issue with 3D generation models like TRELLIS is that they often fail to produce high-quality results for inputs outside the domain of their 3D training datasets, which mostly consist of synthetic objects. \MethodName{} effectively enhances these results, as shown in the figure, producing superior 3D models.

\begin{table}[t!]
    \centering
    \caption{
    \textbf{Quantitative Comparison on 2D Image Refinement}.
    The full-reference metrics are evaluated against the high-quality source images.
    Baseline (LQ) denotes the degraded version of the source images.
    }
    \resizebox{\columnwidth}{!}{%
        \begin{tabular}{lccc|cccc}
            \toprule[1.2pt]
            \multirow{2}{*}{Model} & \multicolumn{3}{c|}{Full-Reference} & \multicolumn{4}{c}{No-Reference} \\
             & PSNR $\uparrow$ & SSIM $\uparrow$ & LPIPS $\downarrow$ & MUSIQ $\uparrow$ & QAlign $\uparrow$ & LIQE $\uparrow$ & TOPIQ $\uparrow$ \\ 
            \midrule[1.0pt]
            Baseline (LQ)              & \textbf{20.701} & \textbf{0.521} & 0.662 & 21.918 & 2.018 & 1.303 & 0.159 \\
            SDEdit (strength = 0.4)               & 19.214 & 0.473 & 0.679 & 22.863 & 2.237 & 1.261 & 0.174 \\
            SDEdit (strength = 0.8)               & 15.255 & 0.379 & 0.746 & 29.190 & 2.860 & 1.321 & 0.214 \\
            NC-SDEdit                  & 17.737 & 0.442 & 0.697 & 25.257 & 2.476 & 1.329 & 0.184 \\
            \RefineMethodName{} (Ours)  & 15.588 & 0.391 &\textbf{ 0.598} & \textbf{39.519} & \textbf{3.337} & \textbf{2.105} & \textbf{0.283} \\
            \bottomrule[1.2pt]
        \end{tabular}%
    }
    \label{table:2d_sdedit_cmp}
\end{table}

\subsection{Evaluation of \RefineMethodName{}}
We analyze \RefineMethodName{} in the image enhancement task using the validation set of LSDIR~\cite{LSDIR}, a large-scale image restoration dataset. From LSDIR's high-quality images, we create low-quality images by downsampling and upsampling them by a factor of 8. For text prompts, we use ChatGPT Vision~\cite{gpt4} to generate descriptions for the images.

\paragraph{Impact of Low-Frequency Component in Diffusion Sampling}
We validate our intuition that the domain information is encoded in the low-frequency component in the latent representation.
To this end, we conduct an experiment as follows.
We prepare a low-quality reference image.
Then, we sample three different images from the same pure Gaussian noise using different strategies.
We sample the first sample following the conventional diffusion process.
We sample the second sample similarly, but we replace the low-frequency component of its latent representation with that of the low-quality reference. To this end, we modify \cref{eq:hfs2} as $z'_t=G_\sigma*\tilde{z}_r + (\delta-G_\sigma)*z_t$.
Finally, for the third image, we replace its high-frequency component with that of the low-quality reference.
We swap the frequency components only for the first four denoising timesteps for both images when low-frequency features primarily emerge.

\begin{figure}[t!]    
\begin{center}
\includegraphics[width=\linewidth]{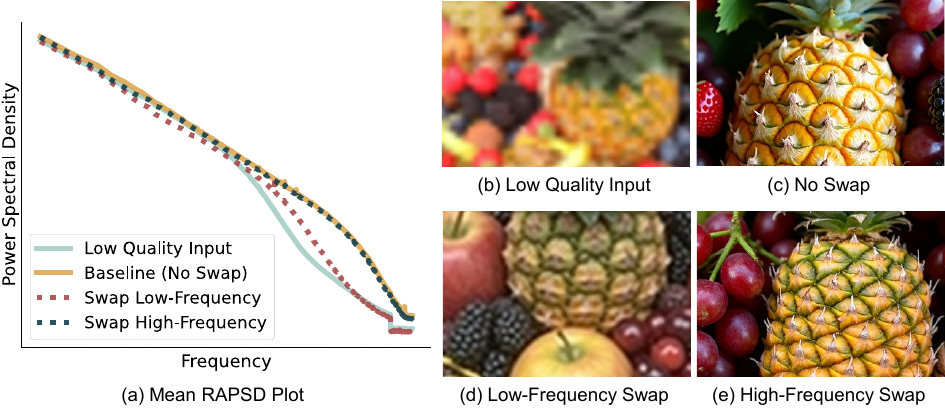}
\end{center}
\caption{
\textbf{Impact of Low-frequency Component in Diffusion Sampling.}
The mean radially averaged power spectral density (RAPSD) graphs of different example images shown in (b)-(e) are shown in (a).
Image (b): ©patrick janicek/flickr.}
\label{fig:rapsd_comparison}
\end{figure}

\cref{fig:rapsd_comparison} compares the three images with their mean radially averaged power spectral density (RAPSD) graphs.
As shown in the figure, when the low-frequency component is replaced with that of the low-quality reference, the diffusion model struggles to synthesize high-frequency details, indicating a shift in the generation path toward the low-quality domain.
Conversely, when only high-frequency component is swapped, the model still produces detailed textures regardless of the reference image's quality.
This observation clearly indicates that the domain information is not in the high-frequency component but in the low-frequency component.

\paragraph{Image Refinement Comparisons}
We validate the effectiveness of \RefineMethodName{} by comparing it with SDEdit and NC-SDEdit~\cite{nc_sdedit}. NC-SDEdit is a video enhancement approach that updates the low-frequency component of latents to match reference frames during the diffusion denoising process, enhancing fidelity. For comparison, we apply NC-SDEdit to single images.
We evaluate the fidelity of the refinement results using full-reference metrics against the high-quality source images: PSNR, SSIM, and LPIPS~\cite{LPIPS}. For quality assessment, we employ no-reference metrics: MUSIQ~\cite{MUSIQ}, LIQE~\cite{LIQE}, TOPIQ~\cite{TOPIQ}, and Q-Align~\cite{QAlign}. 

\begin{figure}[t!]    
\begin{center}
\includegraphics[width=\linewidth,trim=0cm 0cm 0cm 0cm]{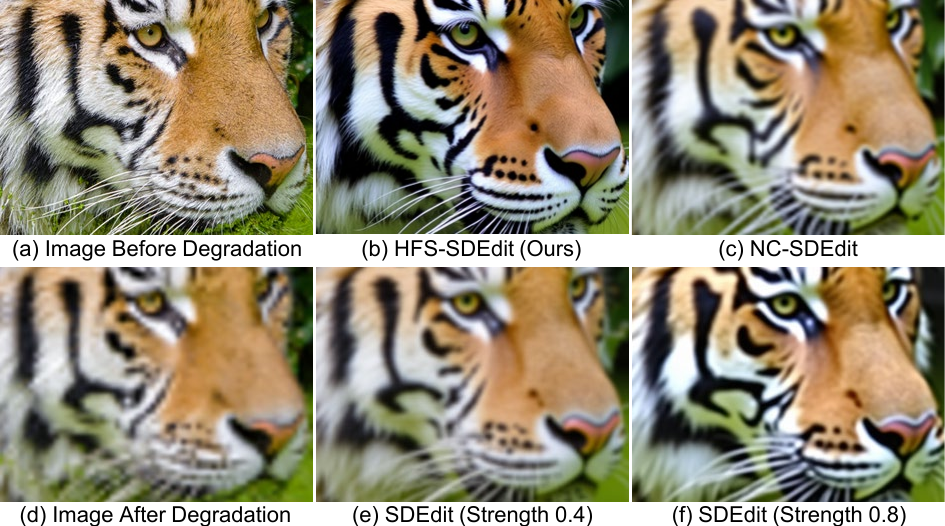}
\end{center}
\caption{
\textbf{Qualitative Comparison on 2D Image Refinement.}
The refinement results in (b), (c), (e), and (f) are obtained from the low-quality image in (d), which was degraded from the image in (a). Image (a): ©Mathias Appel/flickr.
}
\label{fig:sdedit_comp}
\end{figure}

\cref{table:2d_sdedit_cmp} shows the quantitative comparison, where \RefineMethodName{} achieves the best performance in no-reference metrics, demonstrating its capability to produce high-quality outputs. It also obtains the best LPIPS score among its competitors, indicating that the resulting images preserve a high degree of perceptual similarity to the original images. However, because \RefineMethodName{} uses a generative approach to refine the original input, it does not achieve the best PSNR and SSIM scores.
This is common in generative-refinement methods, which prioritize plausible refinement over exact pixel-level fidelity~\cite{supir, psnr_1, psnr2, psnr3}.
Despite this, \cref{fig:sdedit_comp} illustrates that \RefineMethodName{} still produces images with convincing fidelity and high-frequency details.

When examining SDEdit across various strengths, we observe a fidelity-quality trade-off. Lower strength values yield better full-reference metrics but lower no-reference scores, indicating that fidelity is achieved at the expense of quality. Conversely, higher strength values improve perceived quality at the expense of fidelity. Meanwhile, NC-SDEdit’s low-frequency retention condition inadvertently carries over low-quality domain information, leading to subpar results. Consequently, its outputs exhibit both lower fidelity and perceptual quality compared to those of \RefineMethodName{}, reinforcing the advantage of our method's high-frequency-based guidance.

\subsection{Ablation Studies}
\begin{figure}[t!]    
\begin{center}
\includegraphics[width=\linewidth]{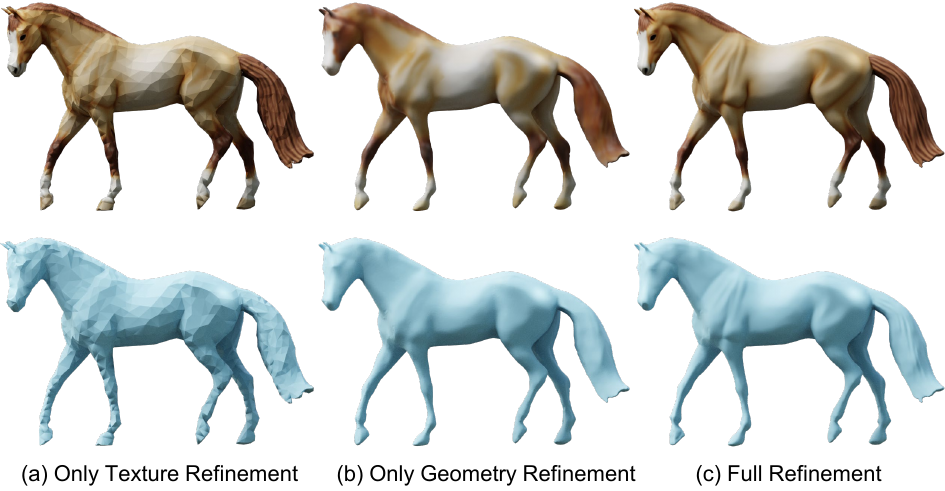}
\end{center}
\caption{
\textbf{Effect of Texture and Geometry Refinement Stages.}
Top: textured meshes, bottom: geometries.
All results are rendered using flat shading.
Input for the experiment: the GSO dataset~\cite{GSO}
}
\label{fig:ablation_geo_tex}
\end{figure}

Finally, we conduct ablation studies to justify our design choices. To demonstrate the necessity of texture and geometry refinement stages, we compare scenarios where only texture or geometry refinement is performed, as shown in \cref{fig:ablation_geo_tex}. Without geometry refinement, the resulting 3D model retains the crude geometry of the input model (\cref{fig:ablation_geo_tex}-a). Conversely, without texture refinement, the geometry refinement stage relies on the low-quality input texture, resulting in minimal geometry improvement (\cref{fig:ablation_geo_tex}-b). Employing both texture and geometry refinement stages yields high-quality textures and geometry that are consistent with each other (\cref{fig:ablation_geo_tex}-c).

\begin{figure}[t!]    
\begin{center}
\includegraphics[width=\linewidth]{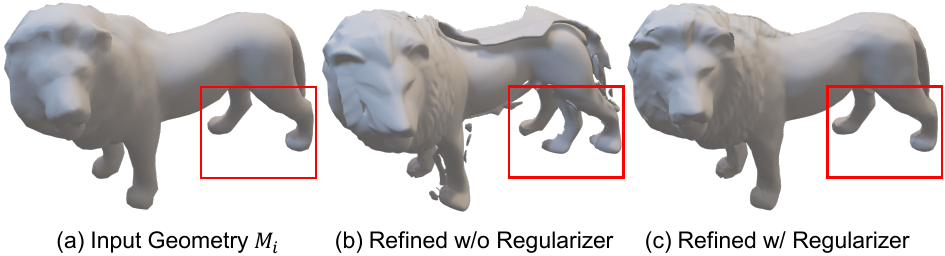}
\end{center}
\caption{
\textbf{Effect of Regularized Normal Integration.}
(a) Initial geometry.
(b) Geometry refinement using normal integration w/o regularization.
(c) Geometry refinement using normal integration w/ regularization (Ours).
Input for the experiment: the GSO dataset~\cite{GSO}
}
\label{fig:ablation_depth_reg}
\end{figure}

\cref{fig:ablation_depth_reg} illustrates the effect of the regularized normal integration. As discussed in \cref{subsec:geo_refine}, normals predicted solely from a refined texture may be inconsistent with the input geometry (\cref{fig:ablation_depth_reg}-a), thus directly using them for geometry refinement causes severe distortions (\cref{fig:ablation_depth_reg}-b). Our regularized normal integration effectively addresses this issue, resulting in high-quality geometry (\cref{fig:ablation_depth_reg}-c).

\section{Conclusion and Future Work}
In this work, we proposed \MethodName{}, a novel 3D model refinement framework that alternates between texture and geometry refinement in a view-by-view fashion to produce high-quality 3D models with well-aligned texture and geometry. We introduced \RefineMethodName{} for texture refinement, leveraging high-frequency guidance to achieve high-fidelity enhancements while mitigating the limitations of previous SDEdit-based methods. Through comprehensive experiments, we demonstrated that our framework achieves state-of-the-art quality refinement of 3D models compared to recent competitors.

\paragraph{Limitations and Future Work}

While our framework produces high-quality textured meshes, it shares a common limitation with similar diffusion-based texturing methods~\cite{Text2Tex, TEXTure, intex}: refinement time increases proportionally with the number of views that need to be generated by the diffusion model. Recent advances in increasing the efficiency of diffusion models~\cite{sd3_turbo, 10.1007/978-981-96-0917-8_9_id-compression} offer potential reductions in our framework's computational cost. Future work will explore integrating such models to optimize \MethodName{}'s processing time while maintaining its high-quality output.

\begin{acks}
This work was supported by Pebblous Inc., and the Institute of Information \& Communications Technology Planning \& Evaluation (IITP) grants (RS-2019-II91906, Artificial Intelligences Graduate School Program (POSTECH), RS-2024-00457882, AI Research Hub Project) funded by the Korea government (MSIT).
\end{acks}

\bibliographystyle{ACM-Reference-Format}
\bibliography{main}

\begin{figure*}[t!]    
\begin{center}
\includegraphics[width=\linewidth]{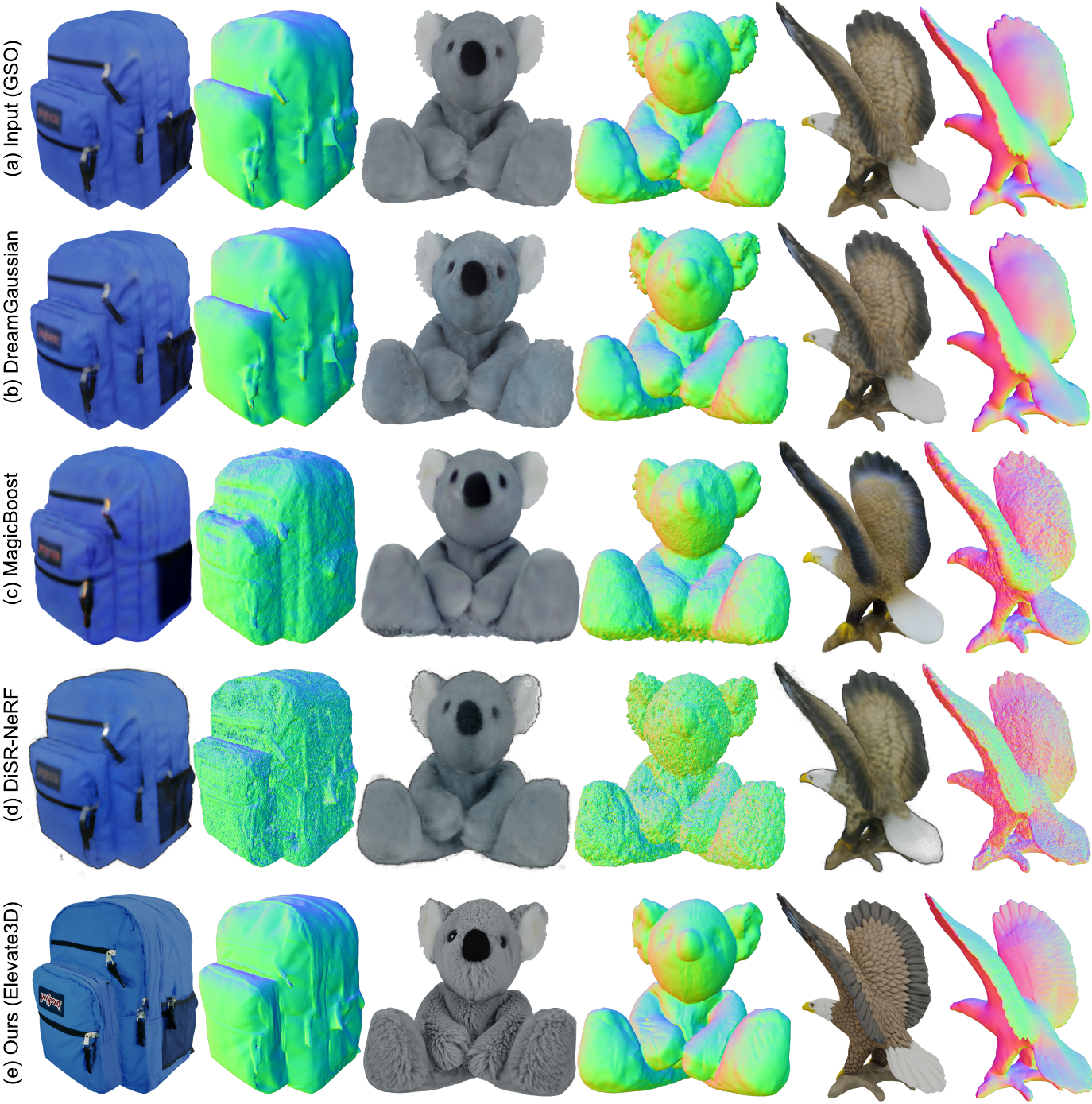}
\end{center}
\caption{
\textbf{Qualitative Comparison on 3D Refinement.}
We compare the 3D refinement results from a low-quality degraded input shown in (a). DreamGaussian~\cite{dreamgaussian} refines only the texture, leaving the geometry degraded as seen in (b). MagicBoost lacks a fidelity constraint, resulting in large deviations from the input as seen in (c). DiSR-Nerf maintains high fidelity but struggles to generate high-frequency details as seen in (d). In contrast, our method effectively refines both texture and geometry while preserving input fidelity while producing high-quality, well-aligned textures and geometries with high quality as shown in (e). Inputs: the GSO dataset~\cite{GSO}
}
\label{fig:qual_comp}
\end{figure*}

\begin{figure*}[t!]    
\begin{center}
\includegraphics[width=\linewidth, trim=0cm 1cm 0cm 0cm]{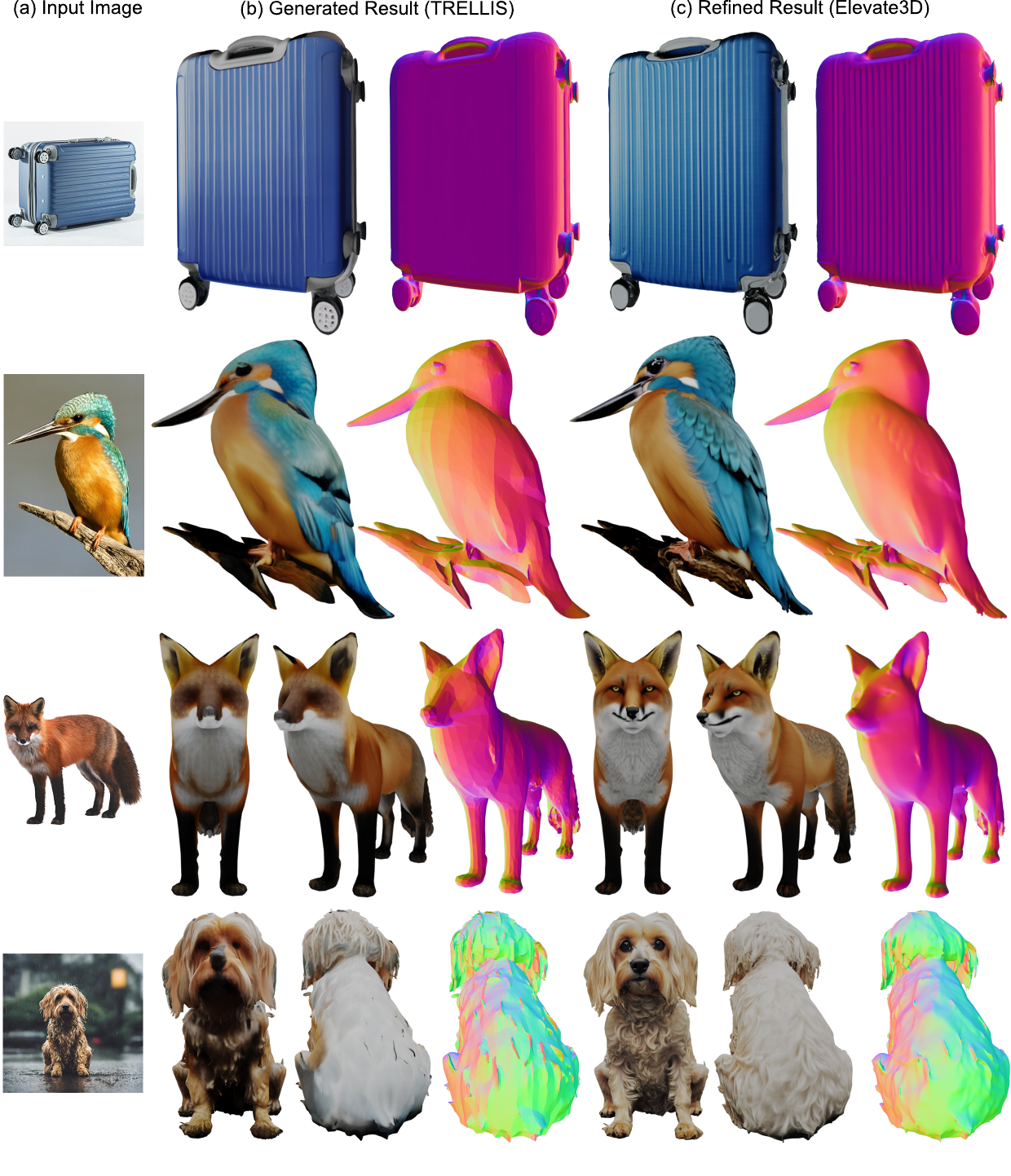}
\end{center}
\caption{
\textbf{Qualitative Results on Refining TRELLIS Outputs.} Due to the domain gap between synthetic training data and real-world images, TRELLIS often struggles to generate high-quality results from real-world inputs images such as in (a), as shown in (b). Therefore, we apply \MethodName{} to refine TRELLIS's outputs. As seen in (c), our method produces realistic textures and accurate geometry, resulting in high-quality refinements. Inputs: ©mec4411/pixabay, ©vimleshtailor/pixabay, ©maja7777/pixabay, ©jacksonmoccelin/pixabay
}
\label{fig:qual_trellis}
\end{figure*}

\FloatBarrier 
\clearpage          %
\appendix           %

\pagenumbering{roman}
\setcounter{page}{1}

\renewcommand{\thetable}{S\arabic{table}}
\setcounter{table}{0}
\renewcommand{\thefigure}{S\arabic{figure}}
\setcounter{figure}{0}

\section{Additional Technical Details}
\label{sec:Additional Technical Details}

\subsection{Details on Geometry Refinement and Filtering}
As described in ~\cref{subsec:geo_refine} of the main paper, the geometry refinement stage aims to find a refined depth map \( z(u,v) \) for the surface patch \( S_i \) by minimizing the following energy functional:
\begin{equation}\label{eq:energy_supp}
\begin{split}
E(z) = \iint \Bigl[\,
&\Bigl(\frac{\partial z}{\partial u} + \frac{n_x(u,v)}{n_z(u,v)}\Bigr)^2
+ \Bigl(\frac{\partial z}{\partial v} + \frac{n_y(u,v)}{n_z(u Vv)}\Bigr)^2
\Bigr] \,du\,dv \\
&\quad + \lambda \iint \Bigl(z(u,v) - d(u,v)\Bigr)^2 \,du\,dv.
\end{split}
\end{equation}
Here, the first term enforces consistency with the estimated normal map \( \mathbf{n}_i = [n_x, n_y, n_z]^\top \) and the second term regularizes the solution towards the depth \( d(u,v) \) rendered from the existing mesh \( M_i \). 
We adapt the normal integration method proposed by Cao et al.~\shortcite{bini} to perform this minimization. 
Their core contribution is a bilaterally weighted functional designed to handle potential depth discontinuities inherent in surfaces estimated from normal maps. Instead of assuming a globally smooth surface, their method operates under the \emph{semi-smooth surface assumption}, allowing for one-sided discontinuities. They introduce bilateral weights \( w_u(u,v) \) and \( w_v(u,v) \) at each pixel \( (u,v) \) during optimization. These weights reflect the local surface continuity; values close to 0.5 indicate local smoothness in the respective direction (horizontal for \( w_u \), vertical for \( w_v \)), while values approaching 0 or 1 suggest a likely discontinuity boundary.

\paragraph{Filtering of Unreliable Depth Estimate}
The depth map \( z(u,v) \), obtained by minimizing \cref{eq:energy_supp}, represents the refined geometry \( S_i \). However, areas near depth discontinuities, indicated by \( w_u \) or \( w_v \) deviating from 0.5, can lead to unreliable depth estimates in \( z(u,v) \). Directly using these unreliable values during the Poisson surface reconstruction could introduce geometric artifacts when merging \( S_i \) with the mesh \( M_i \).

To mitigate this, we filter the depth map \( z(u,v) \) based on the continuity weights \( w_u \) and \( w_v \) computed during the optimization process. We identify pixels \( (u,v) \) where the estimated surface is potentially unreliable or discontinuous by checking if either weight significantly deviates from the ideal smooth value of 0.5. We mark a pixel as unreliable if:
\begin{equation}
    \begin{aligned}
    w_u(u,v) < 0.4 \quad &\text{or} \quad w_u(u,v) > 0.6,\qquad\text{or}\\
    \quad w_v(u,v) < 0.4 \quad &\text{or} \quad w_v(u,v) > 0.6.
    \label{eq:depth_filtering}
    \end{aligned}
\end{equation}

A binary mask is then created, keeping only those pixels where \emph{both} \( w_u(u,v) \) and \( w_v(u,v) \) fall within the confidence interval [0.4, 0.6]. This mask highlights regions deemed to be reliably estimated and locally smooth. To further ensure robustness and remove potentially isolated unreliable pixels or thin artifacts near discontinuity boundaries, this binary mask is processed with morphological erosion using a \( 3 \times 3 \) kernel. The final eroded mask defines the reliable region of the depth map \( z(u,v) \) that is subsequently used in the Poisson surface reconstruction step to update the mesh \( M_i \), ensuring a cleaner and more robust integration of the refined geometry. The seams visible in \cref{fig:method_geometry_refine}-b of the main paper are a direct result of this filtering process of removing pixels around detected discontinuities.

\subsection{Projection Mapping Implementation Details}

As mentioned in  ~\cref{subsec:geo_refine} of the main paper, after geometry refinement yields the updated mesh \( \tilde{M}_i \), we project the corresponding refined texture image \( I'_i \) along with other relevant textures onto \( \tilde{M}_i \) to produce the textured mesh \( M_{i+1} \). This projection mapping is implemented via a custom OpenGL fragment shader. This shader requires several pre-computed data structures passed as uniforms to operate. These include arrays storing all source texture images (\texttt{Textures[]}) corresponding to the camera path views $\mathcal{V}=\{v_0, \ldots, v_k\}$, alongside a status array (\texttt{IsRefined[]}) indicating which textures have been refined by \RefineMethodName{}. Additionally, parameters defining the projection for each view are needed: the projection direction (\texttt{ProjDirections[]}) and the orthographic view and projection matrices (\texttt{ProjViewMats[]}, \texttt{ProjProjMats[]}), derived from the camera's pose for that view. Finally, pre-rendered depth maps (\texttt{DepthMaps[]}) from each viewpoint are crucial for handling occlusions during the blending process. With these inputs prepared, the fragment shader executes the logic outlined in ~\cref{alg:projection_mapping} for each surface fragment.

\begin{algorithm*}[t!]
    \caption{Projection Mapping Fragment Shader Logic}
    \label{alg:projection_mapping}
    \DontPrintSemicolon
    \KwIn{fragPosition, fragNormal \tcp{Fragment attributes}}
    \KwData{Textures[], DepthMaps[], NumTextures, IsRefined[] \tcp{Texture data}}
    \KwData{ProjDirections[], ProjViewMats[], ProjProjMats[] \tcp{Projection parameters}}
    \KwData{Epsilon \tcp{Depth tolerance}}
    \KwOut{finalColor \tcp{Output fragment color}}
    \BlankLine %

    accumColor $\leftarrow$ (0, 0, 0)\;
    totalWeight $\leftarrow$ 0.0\;
    normFragNormal $\leftarrow$ normalize(fragNormal)\;
    \BlankLine

    \For{$j \leftarrow 0$ \KwTo NumTextures $- 1$}{
        \tcp{Initialize weight for texture j}
        weight $\leftarrow$ 1.0\; 
        
        \tcp{Calculate view-dependent alignment weight}
        alignment $\leftarrow$ dot(normFragNormal, ProjDirections[j])\;
        alignmentWeight $\leftarrow$ smoothstep(0.3, 1.0, clamp(alignment, 0.0, 1.0))\;
        weight $\leftarrow$ weight $\times$ alignmentWeight\;
        \BlankLine

        \tcp{Modulate weight based on refinement status}
        \uIf{IsRefined[j]}{
            \tcp{Keep weight if refined}
            weight $\leftarrow$ weight 
        }
        \Else{
            \tcp{Down-weight if unrefined}
            weight $\leftarrow$ weight $\times$ 1e-8\; 
        }
        \BlankLine

        \tcp{Project fragment and get texture coordinates}
        fragTexUV $\leftarrow$ project(fragPosition, ProjViewMats[j], ProjProjMats[j])\; 
        \BlankLine

        \tcp{Perform occlusion test using depth maps}
        isOccluded $\leftarrow$ checkOcclusion(fragPosition, fragTexUV, DepthMaps[j], ProjViewMats[j], ProjProjMats[j], Epsilon)\;
        \If{isOccluded}{
            weight $\leftarrow$ 0.0\;
        }
        \BlankLine

        \tcp{Sample texture and check transparency}
        texColor $\leftarrow$ sampleTexture(Textures[j], fragTexUV)\;
        \If{texColor.alpha < 1.0}{
            \tcp{This case indicates backround in our implementation}
            weight $\leftarrow$ 0.0\;
        }
        \BlankLine

        \tcp{Accumulate weighted color}
        \If{weight > 0}{
             accumColor $\leftarrow$ accumColor + texColor.rgb $\times$ weight\;
             totalWeight $\leftarrow$ totalWeight + weight\;
        }
    } %
    \BlankLine

    \uIf{totalWeight > 1e-8}{
        finalColor.rgb $\leftarrow$ accumColor / totalWeight\;
        finalColor.alpha $\leftarrow$ 1.0\;
    }
    \Else{
        \tcp{Default color (e.g., black)}
        finalColor $\leftarrow$ (0, 0, 0, 0)\; 
    }
    \BlankLine
    \KwRet{finalColor}\; %
\end{algorithm*}

The core of the blending logic within  ~\cref{alg:projection_mapping} lies in calculating an appropriate weight (\texttt{weight}) for each texture's potential contribution. This weighting is carefully designed to ensure high-quality results:
\begin{itemize}
    \item \textbf{View-dependent Alignment:} The \texttt{smoothstep(0.3, 1.0, ...)} function applied to the cosine similarity between the surface normal and projection direction ensures that views nearly perpendicular to the surface contribute strongly, while contributions smoothly fall off to zero for views at grazing angles (beyond approximately 72.5\(^{\circ}\)), preventing artifacts from oblique projections.
    \item \textbf{Refinement Priority:} By drastically reducing the weight of unrefined textures (\( \times 10^{-8} \)) compared to refined ones (\( \times 1.0 \)), the algorithm ensures that the high-quality details introduced by \RefineMethodName{} are preferentially used in the final texture wherever a refined view provides relevant, visible information.
    \item \textbf{Occlusion and Transparency:} Setting the weight to zero for occluded fragments (based on depth map comparison using the conceptual \texttt{checkOcclusion} function) or for fragments projecting onto transparent background regions of the source textures (via the conceptual \texttt{sampleTexture} function checking alpha) prevents projecting incorrect colors or background details onto the foreground mesh surface.
\end{itemize}
This combination of projection (represented conceptually by the \texttt{project} function), visibility checks, and carefully designed weighting allows the shader to synthesize a seamless, UV-free texture on the final mesh, integrating the best available information from multiple viewpoints.

\subsection{Further Technical Details} 
\paragraph{Texture Refinement}
To refine the texture at a target view \( v_i \), we adopt several techniques from the state-of-the-art mesh texturing method Paint3D~\cite{paint3d} in the texture refinement process with \RefineMethodName{}. Specifically, we implement a multi-view depth-aware texture sampling strategy by horizontally concatenating the renderings from the initial viewpoint \( v_0 \) and the target viewpoint \( v_i \), where the camera path is defined as \( \mathcal{V} = \{v_0, \ldots, v_k\} \).
We render depth, RGB, and refinement mask images, resulting in corresponding depth, RGB, and mask grid images.
In the mask grid, only the right half is used as the refinement mask.
Subsequently, we perform multi-view depth-aware texture refinement with \RefineMethodName{}. 
For the initial viewpoint \( v_0 \), we render the next planned view for refinement but utilize only the refined image corresponding to \( v_0 \). For depth conditioning, we employ a variant of FLUX\footnote{\url{https://huggingface.co/black-forest-labs/FLUX.1-Depth-dev}}.

\paragraph{Refinement View Selection}
Inspired by the automatic camera selection scheme introduced in the mesh texturing literature, Text2Tex~\cite{Text2Tex}, we choose the camera path $\mathcal{V} = \left\{v_0, \ldots, v_n, \ldots, v_k\right\}$ as follows.
We begin by defining a sparse camera schedule with polar angles $\left[45^\circ, 90^\circ, 135^\circ\right]$ and corresponding azimuthal angles $\left[0^\circ, 45^\circ, 180^\circ, 270^\circ\right]$. Since an orthographic camera is used, these viewpoints cover most of the visible regions for refinement. However, we employ an automatic camera selection process to address areas that remain unrefined.
We sample 100 views from the sphere. For each viewpoint $v_n$, we render the object to obtain a foreground mask $m_{\text{fg}}$ and the refinement mask $m$. We also compute a cosine similarity map $m_{\text{cos}}$ to evaluate viewpoint quality. We then calculate the ratio $r_n$ as
$$ r_n =  \frac{m_i \odot {m_{\text{cos}}}}{m_{\text{fg}}}$$
for each viewpoint and select the one with the highest ratio for further refinement, ensuring we prioritize the most informative angle. Finally, when the ratio across the object falls below $0.02$, we conclude that sufficient coverage has been achieved and terminate the refinement process.

\paragraph{Geometry Refinement Detail}
~\cref{subsec:geo_refine} of the main paper describes inferring the normal map \(\mathbf{n}_i\) from the refined texture \(I'_i\). In practice, our implementation incorporates an additional normal blending step prior to normal integration to further enhance the robustness of the geometry refinement stage. 
Specifically, before minimizing the energy functional, we blend two normal maps: the map \(\mathbf{n}_i\), inferred from \(I'_i\) that captures fine, texture-consistent details, and a base normal map representing the current geometry of the mesh \(M_i\).
This blending, performed using the UDN blending method~\cite{udn}, helps combine the details from the refined texture's normal map with the underlying structure of the existing mesh geometry. 

\section{Additional Experiments}
\label{sec:Additional Experiments}

\begin{table}[t!]
    \centering
    \caption{\textbf{Parameter Sweep of $\sigma$ in $G_\sigma$ and replacement threshold $t_{\text{stop}}$}.
    Starting the denoising process from $T=30$, the swapping is performed until $t_{\text{stop}}$. The table presents the refinement results according to various combinations of $\sigma$ and $t_{\text{stop}}$ values. Full-reference metrics include PSNR, SSIM, and LPIPS, while non-reference metrics include MUSIQ, QAlign, LIQE, and TOPIQ.
    }
    \resizebox{\columnwidth}{!}{%
    \begin{tabular}{c|c|c c c|c c c c}
        \toprule[1.2pt]
        \multirow{2}{*}{$\sigma$} & \multirow{2}{*}{$t_{\text{stop}}$} & \multicolumn{3}{c|}{Full-Reference Metrics} & \multicolumn{4}{c}{Non-Reference Metrics} \\
        & & PSNR $\uparrow$  & SSIM $\uparrow$  & LPIPS $\downarrow$ & MUSIQ $\uparrow$ & QAlign $\uparrow$ & LIQE $\uparrow$ & TOPIQ $\uparrow$\\ 
        \midrule
        \multirow{4}{*}{2} 
            & 22  & 12.8984 & 0.3144 & 0.6427 & 67.1750 & 4.5764 & 4.0468 & 0.5798\\[1mm]
            & 20 & 13.5825 & 0.3353 & 0.6215 & 58.0347 & 4.1461 & 3.2241 & 0.4679\\[1mm]
            & 18 & 14.2574 & 0.3562 & 0.6136 & 49.3993 & 3.6700 & 2.5849 & 0.3793\\[1mm]
            & 16 & 14.8523 & 0.3739 & 0.6084 & 41.9640 & 3.2286 & 2.0822 & 0.3057\\[1mm]
        \midrule

        \multirow{4}{*}{4} 
            & 22 & 14.1419 & 0.3514 & 0.6020 & 54.7414 & 4.1851 & 3.0206 & 0.4206\\[1mm]
            & 20 & 14.8375 & 0.3702 & 0.5980 & 45.7231 & 3.6959 & 2.4237 & 0.3354\\[1mm]
            & 18 & 15.5876 & 0.3906 & 0.5982 & 39.5193 & 3.3370 & 2.1052 & 0.2828\\[1mm]
            & 16 & 16.2478 & 0.4057 & 0.5969 & 34.8861 & 3.0465 & 1.8206 & 0.2390\\[1mm]
        \midrule

        \multirow{4}{*}{16} 
            & 22 & 16.2883 & 0.4014 & 0.6864 & 31.1365 & 3.0541 & 1.4288 & 0.2341\\[1mm]
            & 20 & 17.0381 & 0.4210 & 0.6840 & 28.2405 & 2.7874 & 1.3988 & 0.2074\\[1mm]
            & 18 & 17.7739 & 0.4390 & 0.6806 & 26.0664 & 2.6046 & 1.3661 & 0.1913\\[1mm]
            & 16 & 18.3762 & 0.4529 & 0.6752 & 25.1298 & 2.4729 & 1.3231 & 0.1845\\[1mm]
            
        \bottomrule[1.2pt]
    \end{tabular}%
    }
    \label{table:ablation_mask_thresh}
\end{table}

\begin{table}[t!]
    \centering
    \caption{
    \textbf{Comparison with ProlificDreamer}.
    \MethodName{} consistently achieves the better scores across various quality metrics, highlighting its high-quality, visually appealing 3D refinement.
    }
    \resizebox{\columnwidth}{!}{%
        \begin{tabular}{lccc|cccc}
            \toprule[1.2pt]
            \multirow{2}{*}{Method} & \multicolumn{3}{c|}{Full-Reference} & \multicolumn{4}{c}{No-Reference} \\
             & PSNR$\uparrow$ & SSIM$\uparrow$ & LPIPS$\downarrow$ & MUSIQ$\uparrow$ & LIQE$\uparrow$ & TOPIQ$\uparrow$ & Q-Align$\uparrow$ \\ 
            \midrule[1.0pt]
            LQ (Baseline)       & \textbf{33.202} & \textbf{0.966} & \textbf{0.057} & 61.241 & 2.069 & 0.457 & 2.690 \\
            Ours                & 26.163 & 0.941 & 0.070 & \textbf{66.527} & \textbf{2.774} & \textbf{0.529} & \textbf{3.215} \\
            ProlificDreamer     & 18.037 & 0.904 & 0.156 & 57.307 & 1.989 & 0.439 & 1.893 \\
            \bottomrule[1.2pt]
        \end{tabular}%
    }
    \label{table:prolific_dreamer}
\end{table}

\begin{table}[t!]
    \centering
    \caption{
    \textbf{Quantitative Results on the Ablation}.
     We compare with the cases where only texture or geometry refinement is performed.
    }
    \resizebox{\columnwidth}{!}{%
        \begin{tabular}{lccc|cccc|c}
            \toprule[1.2pt]
            \multirow{2}{*}{Model} & \multicolumn{3}{c|}{Full-Reference} & \multicolumn{4}{c|}{No-Reference} & \multirow{2}{*}{Normal FID$\downarrow$} \\
             & PSNR$\uparrow$ & SSIM$\uparrow$ & LPIPS$\downarrow$ & MUSIQ$\uparrow$ & LIQE$\uparrow$ & TOPIQ$\uparrow$ & Q-Align$\uparrow$ & \\ 
            \midrule[1.0pt]
            LQ (Baseline)             & \textbf{33.202} & \textbf{0.966} & \textbf{0.057} & 61.241 & 2.069 & 0.457 & 2.690 & 60.786 \\
            FULL                      & 26.163 & 0.941 & 0.070 & 66.527 & 2.774 & 0.529 & 3.215 & 52.195 \\
            Only Geometry Refinement  & 28.365 & 0.954 & 0.068 & 61.086 & 2.032 & 0.465 & 2.599 & \textbf{48.043} \\
            Only Texture Refinement   & 26.393 & 0.939 & 0.067 & \textbf{68.717} & \textbf{3.188} & \textbf{0.572} & \textbf{3.454} & 55.465 \\
            \bottomrule[1.2pt]
        \end{tabular}%
    }
    \label{table:quant_ablation}
\end{table}

\begin{table}[t!]
    \centering
    \caption{
    \textbf{Experimental Results with a Different Diffusion Backbone}.
    The full-reference metrics are evaluated against the high-quality source images. LQ (Baseline) means the low-quality reference images degraded from the high-quality source images.
    }
    \resizebox{\columnwidth}{!}{%
        \begin{tabular}{lccc|cccc}
            \toprule[1.2pt]
            \multirow{2}{*}{Model} & \multicolumn{3}{c|}{Full-Reference} & \multicolumn{4}{c}{No-Reference} \\
             & PSNR$\uparrow$ & SSIM$\uparrow$ & LPIPS$\downarrow$ & MUSIQ$\uparrow$ & QAlign$\uparrow$ & LIQE$\uparrow$ & TOPIQ$\uparrow$ \\ 
            \midrule[1.0pt]
            LQ (Baseline)            & \textbf{20.701} & \textbf{0.521} & 0.662 & 21.918 & 2.018 & 1.303 & 0.159 \\
            SDEdit (strength=0.4)    & 18.982 & 0.457 & 0.689 & 23.090 & 2.204 & 1.316 & 0.155 \\
            SDEdit (strength=0.8)    & 14.809 & 0.372 & 0.741 & 31.714 & 2.854 & 1.446 & 0.224 \\
            NC-SDEdit                & 17.344 & 0.424 & 0.717 & 24.730 & 2.347 & 1.274 & 0.168 \\
            Ours                     & 15.853 & 0.376 & \textbf{0.601} & \textbf{41.754} & \textbf{3.415} & \textbf{1.876} & \textbf{0.295} \\
            \bottomrule[1.2pt]
        \end{tabular}%
    }
    \label{table:backbone_ablation}
\end{table}

\begin{table}[t!]
    \centering
    \caption{
    \textbf{Quantitative Analysis of Fig. 10 in the Main Paper}.
    We compare the quantitative results of the objects shown in Fig. 10 in the Main Paper.
    }
    \resizebox{\columnwidth}{!}{%
        \begin{tabular}{lcccc|cccc}
            \toprule[1.2pt]
            \multirow{2}{*}{Method} & \multicolumn{4}{c|}{Full-Reference} & \multicolumn{4}{c}{No-Reference} \\
             & PSNR$\uparrow$ & SSIM$\uparrow$ & LPIPS$\downarrow$ & CLIP Sim$\uparrow$ & MUSIQ$\uparrow$ & LIQE$\uparrow$ & TOPIQ$\uparrow$ & Q-Align$\uparrow$ \\ 
            \midrule[1.0pt]
            LQ (Baseline)     & \textbf{32.086} & 0.940 & 0.096 & 88.640 & 61.203 & 2.022 & 0.429 & 2.577 \\
            Ours              & 25.342 & 0.902 & 0.098 & \textbf{90.577} & \textbf{67.674} & \textbf{2.756} & \textbf{0.540} & \textbf{3.277} \\
            DreamGaussian     & 31.559 & \textbf{0.942} & \textbf{0.086} & 88.387 & 61.669 & 2.083 & 0.457 & 2.704 \\
            \bottomrule[1.2pt]
        \end{tabular}%
    }
    \label{table:fig10_analysis}
\end{table}

\begin{table}[t!]
    \centering
    \caption{\textbf{Quantitative Comparison on 3D Refinement Using Full-reference metrics}.
    DreamGaussian~\cite{dreamgaussian} refines only textures while other methods refine both texture and geometry.
    }
    \resizebox{\columnwidth}{!}{%
        \begin{tabular}{l c c c c}
            \toprule[1.2pt]
            Model & PSNR$\uparrow$ & SSIM$\uparrow$ & LPIPS$\downarrow$ & CLIP$\uparrow$ \\
            \midrule[1.0pt]
            LQ (Baseline) & \textbf{33.202} & \textbf{0.966} & 0.057 & 90.688 \\
            Ours & 26.163 & 0.941 & 0.070 & 89.886 \\
            DreamGaussian~\cite{dreamgaussian} & 32.720 & 0.965 & \textbf{0.056} & \textbf{90.890} \\
            DiSR-NeRF~\cite{disr} & 30.294 & 0.949 & 0.088 & 88.696 \\
            MagicBoost~\cite{magicboost} & 23.865 & 0.934 & 0.106 & 83.412 \\
            ProlificDreamer~\cite{prolificdreamer} & 18.037 & 0.904 & 0.156 & 78.509 \\
            \bottomrule[1.2pt]
        \end{tabular}
    }
    \label{table:3d_refinement_fullref}
\end{table}

\subsection{The Effect of the Parameters in \RefineMethodName{}}
As discussed in the main paper, \RefineMethodName{} employs a Gaussian kernel $G_\sigma$, and the high-frequency replacement is performed until it reaches the time step $t_{\text{stop}}$. To analyze the effect of different parameter choices for $\sigma$ and $t_{\text{stop}}$ in \RefineMethodName{}, we conduct the same low-quality image refinement experiment presented in the main paper.
For assessing refinement fidelity, we utilize full-reference metrics, including PSNR, SSIM~\cite{SSIM}, and LPIPS~\cite{LPIPS}. To evaluate perceptual quality, we employ non-reference metrics such as MUSIQ~\cite{MUSIQ}, LIQE~\cite{LIQE}, TOPIQ~\cite{TOPIQ}, and Q-Align~\cite{QAlign}.
As shown in ~\cref{table:ablation_mask_thresh}, for a fixed $\sigma$, increasing the number of replacement steps improves fidelity, leading to better full-reference metrics. However, this also increases the risk of preserving degraded details from the low-quality image, lowering the performance on non-reference metrics. Conversely, increasing $\sigma$ allows a broader band of frequencies to be preserved for a fixed number of replacement steps. While this improves fidelity and enhances the full-reference metrics, it also risks incorporating low-quality domain information from the low-quality reference image, thereby negatively impacting non-reference metrics.
The parameter combination of $\sigma = 4$ and $t_{\text{stop}} = 18$, used in the main experiment, is observed to provide a balance, achieving both high-fidelity and high-quality refinement.

\subsection{Comparison with ProlificDreamer}
We provide a comparison between our method and ProlificDreamer~\cite{prolificdreamer}.
While ProlificDreamer was initially  developed for text-to-3D synthesis, it can also be applied to refine existing 3D models due to its VSD-loss-based framework. We use the geometry and texture refinement stages of ProlificDreamer for refinement. As shown in \cref{table:prolific_dreamer}, our method significantly outperforms ProlificDreamer across all evaluated metrics, including MUSIQ~\cite{MUSIQ}, LIQE~\cite{LIQE}, TOPIQ~\cite{TOPIQ}, and Q-Align~\cite{QAlign}.
ProlificDreamer showed lower fidelity refinements due to the lack of explicit fidelity constraints.
Quantitatively, ProlificDreamer exhibits substantially lower PSNR, SSIM~\cite{SSIM}, and higher LPIPS~\cite{LPIPS} values. Qualitatively, we can see in~\cref{fig:supple_gso} that direct VSD application often results in artifacts, such as multi-faced Janus effects, due to the absence of explicit fidelity constraints.

\subsection{Quantitative Ablation Study of Texture and Geometry Refinement}
We performed a quantitative ablation study to evaluate contributions from geometry and texture refinements as shown in \cref{table:quant_ablation}.
For geometry evaluation, we computed the FID between high-quality normal maps and those from each refinement method. The full refinement achieves balanced and competitive results across both geometry and texture metrics, while geometry-only and texture-only refinements perform well in their respective domains but not both.
This highlights our method's holistic improvement across both geometry and texture aspects.

\subsection{Experimental Results with a Different Diffusion Backbone}
To examine the robustness of \RefineMethodName{} with respect to the diffusion backbone, we repeated our main experiment using a less powerful model, Stable Diffusion 3.5 medium~\cite{sd3}, under identical experimental conditions. Results in \cref{table:backbone_ablation} reflect similar trends to the main experiments, demonstrating improvements in non-reference metrics and LPIPS~\cite{LPIPS} scores.

\subsection{Quantitative Analysis of ~\cref{fig:qual_comp} in the Main Paper}
~\cref{fig:qual_comp} in the main paper qualitatively demonstrates substantial visual enhancements achieved by our method over DreamGaussian~\cite{dreamgaussian}. However, the corresponding quantitative analysis using non-reference metrics in~\cref{table:fig10_analysis} reveals less dramatic numerical differences. This highlights a common limitation where such metrics may not fully reflect perceived visual quality improvements. Despite this, the scores do confirm a relative improvement provided by our approach.

\subsection{Quantitative Comparison of 3D Refinement with Full Reference Metrics}
We compare our method with recent 3D model refinement approaches in terms of various full reference metrics: PSNR, SSIM~\cite{SSIM}, LPIPS~\cite{LPIPS}, and CLIP similarity~\cite{CLIP}.
As summarized in \cref{table:3d_refinement_fullref}, we see that when the initial coarse 3D input is already of moderate fidelity, these metrics yield similar scores across different refinement methods. Thus, these metrics may not adequately capture actual perceptual enhancements for the generative refinement task that involves generating details absent in the reference.

\subsection{Robustness to Normal Prediction Failures}
\begin{figure}[t!]    
\begin{center}
\includegraphics[width=\columnwidth]{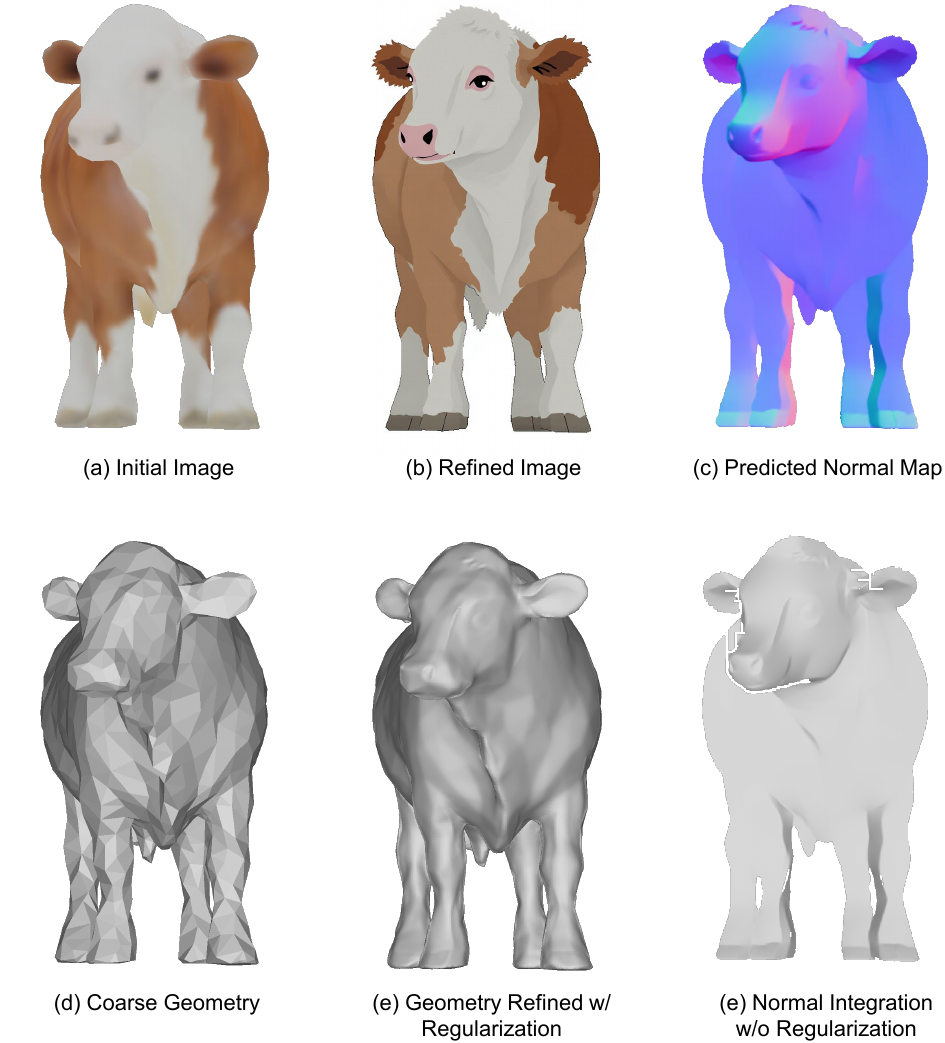}
\end{center}
\vspace{-4.0mm}
\caption{
\textbf{Robustness to Normal Prediction Failure.}
Even when the normal predictor generates poor results, our regularized normal integration successfully preserves the coarse geometry structure, preventing severe distortion.
}
\label{fig:supple_normal_pred_fail}
\end{figure}

After the input image (\cref{fig:supple_normal_pred_fail}-a) is refined with \RefineMethodName{} (\cref{fig:supple_normal_pred_fail}-b), we extract extract geometric cues from the refined image using an off-the-shelf monocular normal estimator~\cite{mari_e2e}.
However, the normal estimator may occasionally produce failure cases. A common failure mode produces overly smooth or detail-less predicted normal maps (\cref{fig:supple_normal_pred_fail}-c). Directly Integrating such a compromised normal map produces flat, featureless surfaces that negate the purpose of refinement (\cref{fig:supple_normal_pred_fail}-e). Our method, however, is robust to such failures due to the regularization term in our energy functional \cref{eq:energy_supp}. This term explicitly encourages the refined surface \(S_i\) to remain close to the depth map derived from the input coarse geometry \(M_i\) (\cref{fig:supple_normal_pred_fail}-d). This allows our method to produce meaningfully refined geometry (\cref{fig:supple_normal_pred_fail}-e) without severe distortions, avoiding catastrophic failure during the refinement iterations.

\subsection{Additional 2D Refinement Results}
\label{sec:Additional Generation Results}
\begin{figure*}[t!]    
\begin{center}
\includegraphics[width=\linewidth]{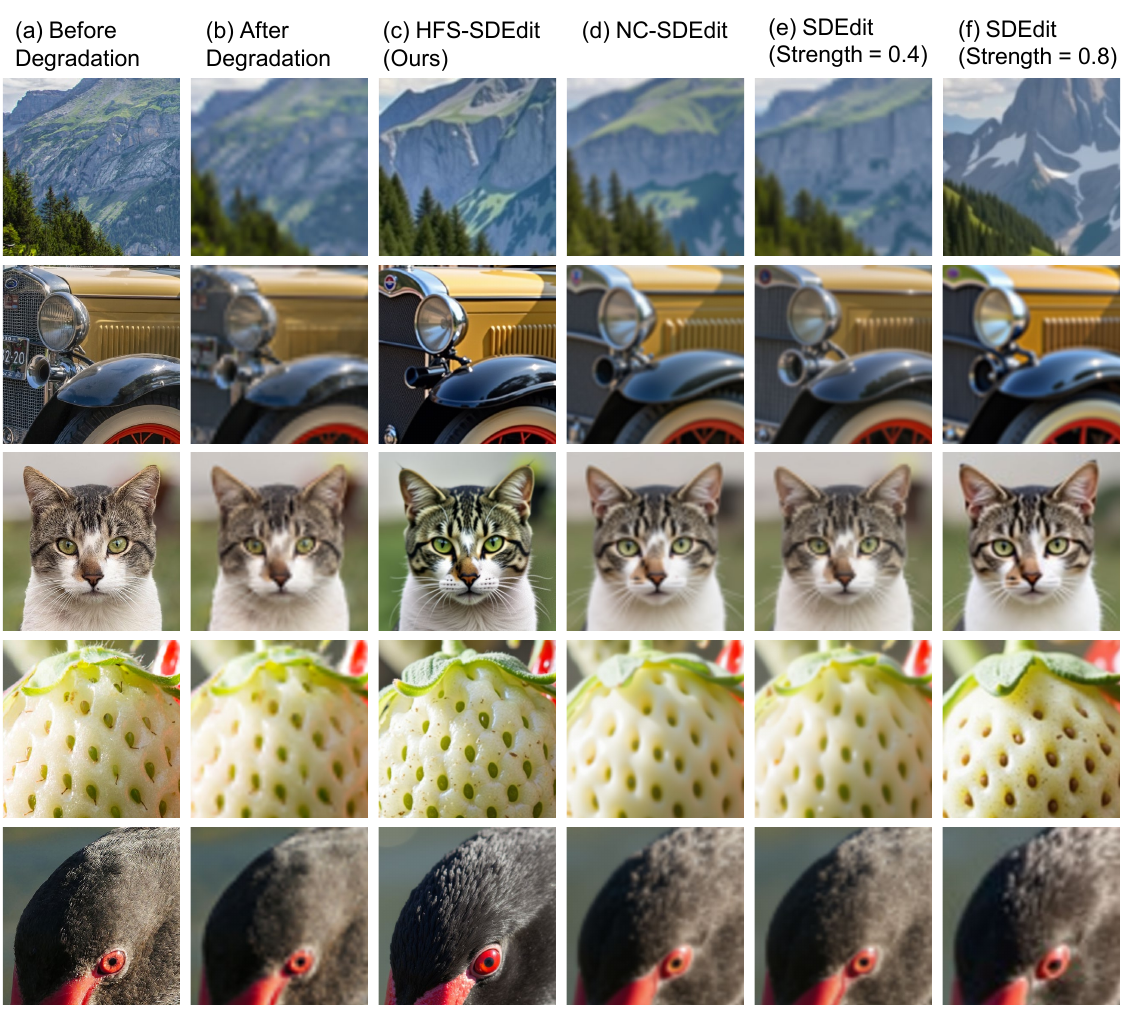}
\end{center}
\vspace{-4.0mm}
\caption{
\textbf{Additional Qualitative Comparison on 2D Image Refinement.}
The refinement results in (c), (d), (e), and (f) are obtained from the low-quality image in (b), which was degraded from the image in (a).
}
\label{fig:supple_sdedit}
\end{figure*}

In~\cref{fig:supple_sdedit}, we show more qualitative comparsion results comparing SDEdit~\cite{sdedit}, NC-SDEdit~\cite{nc_sdedit}, and \RefineMethodName{}. Our method shows a good balance between fidelity and quality.

\subsection{Additional 3D Refinement Results}
\label{sec:Additional Generation Results}
\begin{figure*}[t!]    
\begin{center}
\includegraphics[width=\linewidth]{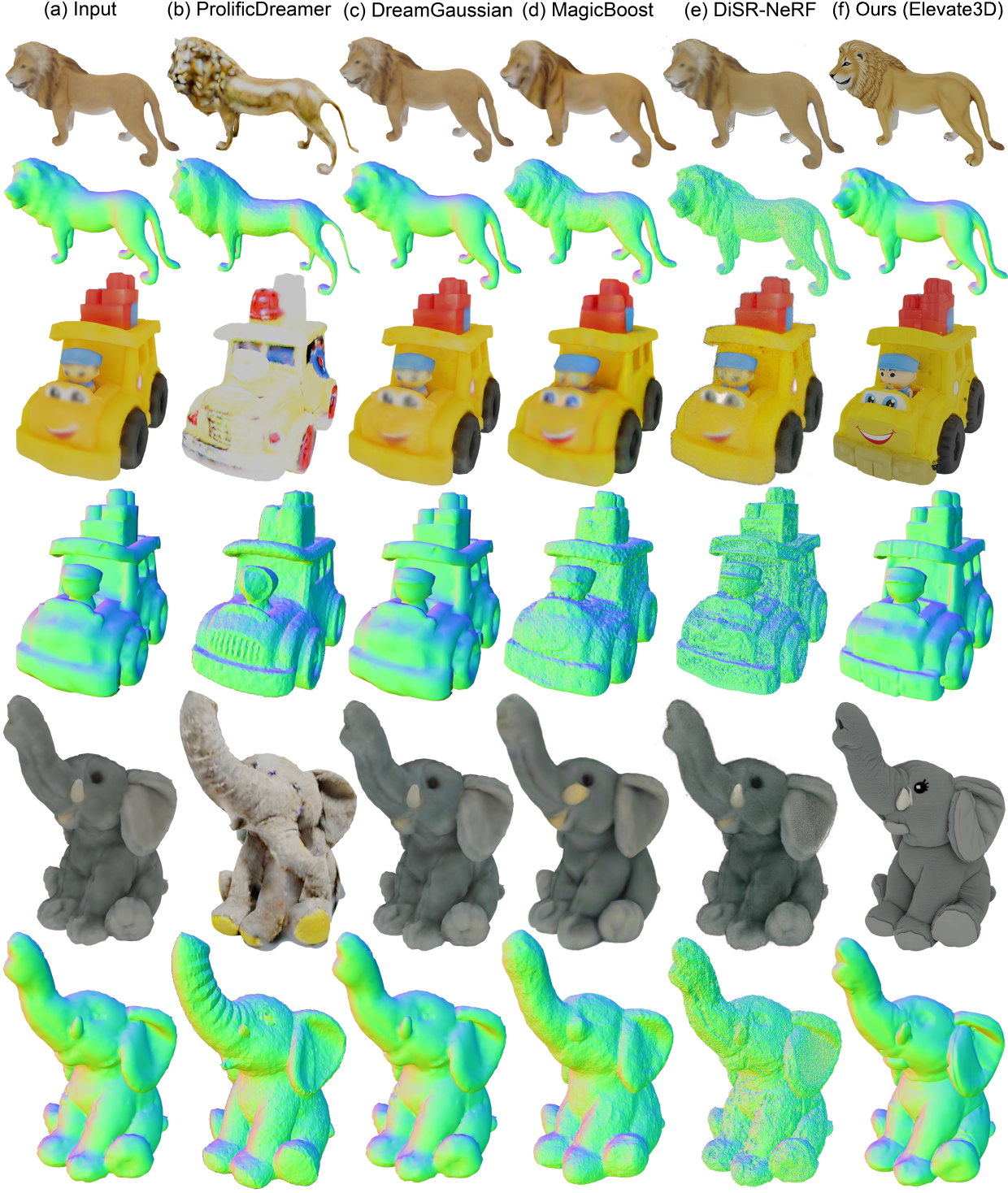}
\end{center}
\vspace{-4.0mm}
\caption{
\textbf{Additional Qualitative Comparison on 3D Refinement}
We show additional 3D refinement comparisons on the degraded GSO dataset. Among all methods, our model produces the highest quality textures with a well-aligned geometry.
}
\label{fig:supple_gso}
\end{figure*}

\begin{figure*}[t!]    
\begin{center}
\includegraphics[width=\linewidth]{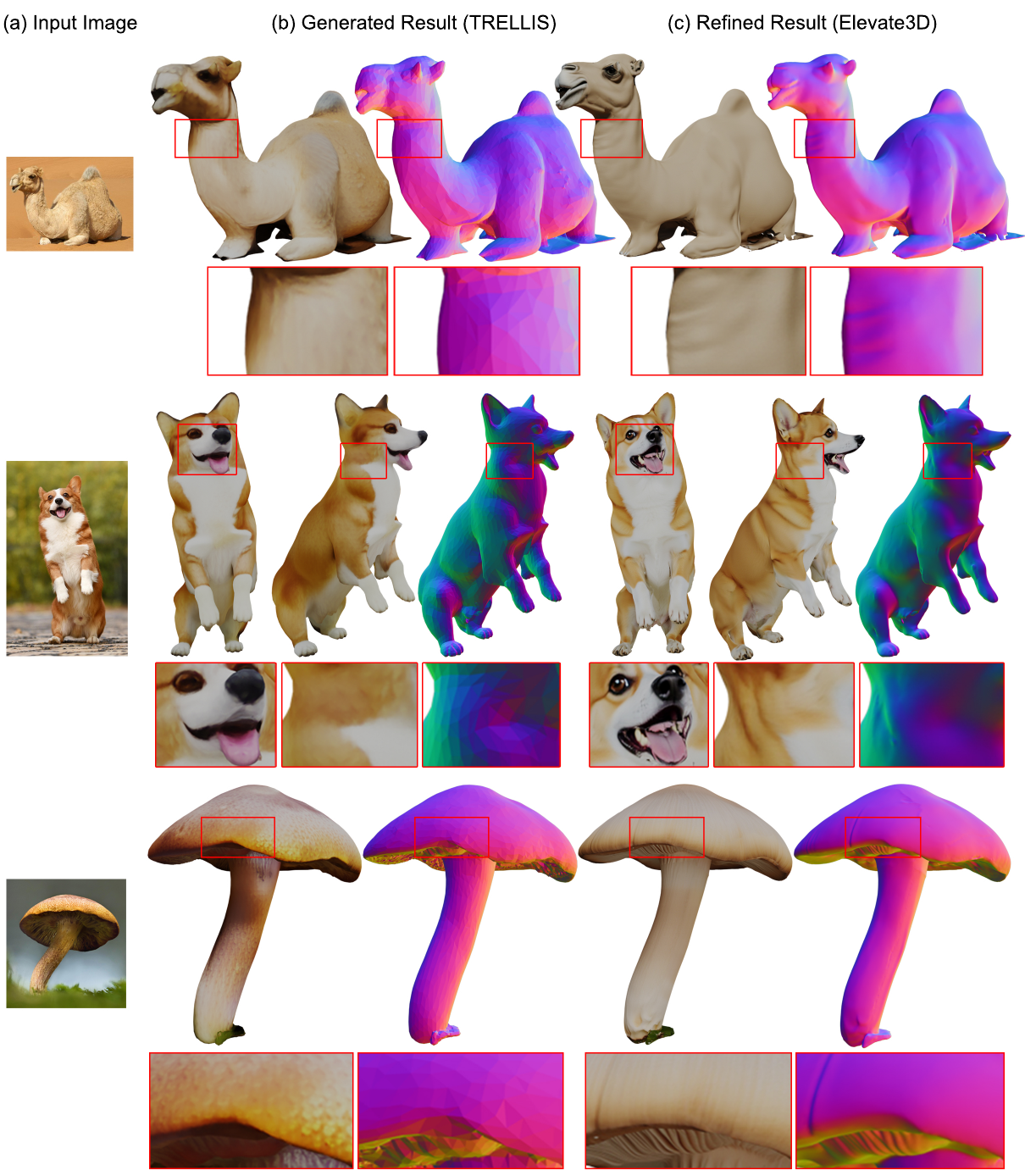}
\end{center}
\vspace{-4.0mm}
\caption{
\textbf{Additional Qualitative Comparison on TRELLIS Outputs.}
We show additional examples of refining 3D generation results of TRELLIS where it generates the 3d in (b) taking real world input image in (a).  As seen in (c), our method produces realistic textures and accurate geometry, resulting in high-quality refinements.
}
\label{fig:supple_trellis}
\end{figure*}

In this section, we present additional qualitative examples. 
In~\cref{fig:supple_gso}, we show more qualitative comparison results for the refinement results of the degraded GSO~\cite{GSO} dataset. In ~\cref{fig:supple_trellis}, we further show more qualitative results for refining the 3D generation results of TRELLIS~\cite{trellis}.
We also provide an additional supplementary video. \textbf{We strongly suggest the readers also see the video} for a better understanding of our model's output quality.

\end{document}